\documentclass[journal]{IEEEtran}

\usepackage{cite}
\usepackage{amsmath,amssymb,amsfonts}
\usepackage{algorithm,algorithmic}
\usepackage{graphicx}
\usepackage{textcomp}
\usepackage{color,xcolor}
\usepackage{bm}
\usepackage{subfigure}
\usepackage{multirow}
\usepackage[normalem]{ulem}
\usepackage{tabularx}
\usepackage{xpatch}
\usepackage{soul}
\usepackage{ulem}
\usepackage{verbatim}
\usepackage{url}
\usepackage{float}

\ifCLASSINFOpdf
\else
\fi

\begin{document}

\title{Location Tracking for Reconfigurable Intelligent Surfaces Aided Vehicle Platoons: Diverse \\ Sparsities Inspired Approaches}

\author{Yuanbin~Chen,~Ying~Wang,~\IEEEmembership{Member,~IEEE,}~Xufeng~Guo,~Zhu~Han,~\IEEEmembership{~Fellow,~IEEE,}\\and~Ping~Zhang,\IEEEmembership{~Fellow,~IEEE}

\thanks{ Manuscript received~31~August~2022; revised 31~March~2023; accepted~3~May~2023. This work was supported by Beijing Natural Science Foundation under Grant 4222011, and in part by NSF CNS-2107216, CNS-2128368, CMMI-2222810, US Department of Transportation, Toyota and Amazon. \textit{(Corresponding author: Ying Wang.)}

Yuanbin Chen, Ying Wang, Xufeng Guo, and Ping Zhang are with the State Key Laboratory of Networking and Switching Technology, Beijing University of Posts and Telecommunications, Beijing 100876, China (e-mail: chen\_yuanbin@163.com; wangying@bupt.edu.cn; brook1711@bupt.edu.cn; pzhang@bupt.edu.cn). 

Zhu Han is with the Department of Electrical and Computer Engineering in the University of Houston, Houston, TX~77004, USA, and also with the Department of Computer Science and Engineering, Kyung Hee University, Seoul, South Korea, 446-701 (e-mail: hanzhu22@gmail.com).
}

}

%



\maketitle

\begin{abstract}
In this paper, we investigate the employment of reconfigurable intelligent surfaces (RISs) into vehicle platoons, functioning in tandem with a base station (BS) in support of the high-precision location tracking. In particular, the use of a RIS imposes additional structured sparsity that, when paired with the initial sparse line-of-sight (LoS) channels of the BS, facilitates beneficial group sparsity. The resultant group sparsity significantly enriches the energies of the original direct-only channel, enabling a greater concentration of the LoS channel energies emanated from the same vehicle location index. Furthermore, the burst sparsity is exposed by representing the non-line-of-sight (NLoS) channels as their sparse copies. This thus constitutes the philosophy of the diverse sparsities of interest. Then, a diverse dynamic layered structured sparsity (DiLuS) framework is customized for capturing different priors for this pair of sparsities, based upon which the location tracking problem is formulated as a maximum a posterior (MAP) estimate of the location. Nevertheless, the tracking issue is highly intractable due to the ill-conditioned sensing matrix, intricately coupled latent variables associated with the BS and RIS, and the spatial-temporal correlations among the vehicle platoon.
To circumvent these hurdles, we propose an efficient algorithm, namely DiLuS enabled spatial-temporal platoon localization (DiLuS-STPL), which incorporates both variational Bayesian inference (VBI) and message passing techniques for recursively achieving parameter updates in a turbo-like way. Finally, we demonstrate through extensive simulation results that the localization relying exclusively upon a BS and a RIS may achieve the comparable  precision performance obtained by the two individual BSs, along with the robustness and superiority of our proposed algorithm as compared to various benchmark~schemes.
\end{abstract}


%
\IEEEpeerreviewmaketitle

\section{Introduction}

With a tidal wave of technological innovation upon us, the sixth-generation (6G) communication system tends to be a game changer in terms of placating the extraordinary promises, particularly in support of enormous amounts of widely heterogeneous information exchange in a hyperfast, low-latency and ultra-reliable manner.  A grand 6G paradigm needs to be materialized by revolution-natured technologies, such as terahertz (THz) communications, reconfigurable intelligent surfaces (RISs), and artificial intelligence, in an effort to advance the existing mechanisms and accommodate the newly encountered particularities \cite{Angeliki,H-V}. 
Reaping plethora of benefits in 6G, connected autonomous vehicles (CAVs) have the potential to provide unparalleled quality of experiences, tremendously improved traffic conditions, road safety, and an abundance of cutting-edge vehicular applications \cite{chen-vtm}. This relies significantly upon a high-definition degree of situational awareness with the physical environment, i.e., the ability to identify its own position. In response, standardization lays the foundation of localization use cases and specifications for sidelinks (e.g., vehicle platoons and coordinated maneuvers), including the recently frozen 3rd generation partnership project (3GPP) Release 17, in an effort to further improve the fifth-generation (5G) localization for fulfilling more stringent criteria, e.g., centimeter-level accuracy~\cite{3GPP-R17,STD-11}. 
The existing vehicle localization technologies, such as global positioning system (GPS), radar, lidar and inertial sensors, however, fall well short of achieving these specifications, thus mandating rethinking ways and new contributing technologies that go beyond the present ones.

The common localization techniques can be informally categorized into indirect approaches and direct thereof. In particular, the indirect localization aims to obtain the intermediate parameters, e.g., angles of arrival (AoA), time of arrival (ToA), and received signal strength (RSS), from the received signals before applying the trick of triangulation to determine user's location. For instance, the data-driven methods including deep learning and fingerprinting are employed in~\cite{AngDom-17-8} and~\cite{FP-CST}, respectively, to estimate the RSS in a massive multiple-input multiple-output (MIMO) system. It is investigated in \cite{AngDom-17-11} that a joint estimate of AoA and ToA is considered for establishing user's location. The direct localization, by contrast, recovers the associated positions straightforwardly from the received signal, which is applicable exclusively in the pure line-of-sight (LoS) environment but can also be extended to the multi-path scenario, based upon which such a geometry-based approach allows the user's location to be constructed by taking full advantage of various LoS and non-line-of-sight (NLoS) components \cite{LOC-5,LOC-6}. Furthermore, the application of the two localization techniques facilitates the location tracking in the context of vehicular networks \cite{LOC-1,LOC-2,LOC-9}, including the geometry-based cooperative localization~\cite{LOC-1}, indirect approach-based tracking~\cite{LOC-2}, and data-driven learning-based method~\cite{LOC-9}. Owing to the high degrees of freedom and large apertures in the array equipped at the base station (BS), the localization accuracy may be improved if used in conjunction with an appropriate LoS/NLoS model that is crafted to appropriately characterize the vehicular multi-path environment. While these initiatives share common facts with the collaborative efforts of multiple BSs, they may, in practice, achieve so at the expense of unacceptable signaling overhead and infrastructure costs.

Benefiting from the digitally controlled meta-surfaces, RIS has been hailed as an impressive technology capable of transforming the conventional uncontrollable wireless channels into their controllable counterparts, thus materializing the concept of ``smart and customizable radio environment"~\cite{RIS-101,Debbah,EC-S}. Intriguingly, a tight amalgamation of RIS with localization is capable of making significant contributions to several challenging issues, with a number of fledgling research efforts already available to date \cite{RIS-LOC-1,reviewer-1, reviewer-2,RIS-LOC-2, RIS-LOC-12, RIS-LOC-13,RIS-LOC-11, RIS-LOC-10}. More explicitly, in \cite{RIS-LOC-1}, the use of a RIS enables the exact construction of distinct RSS values across adjacent locations by appropriately adjusting the phase shifts of the RIS in the presence of only one access point (AP) indoors. It is investigated in \cite{reviewer-1} that a joint localization and synchronization paradigm is conceived based on an optimized design of the BS active precoding and RIS passive beamforming.
A RIS is deployed in \cite{reviewer-2} for investigating the capability of positioning a single-antenna user by taking into account the mobility of the user and spatial-wideband effects in a single-input and single-output (SISO) system.
Furthermore, RIS can be treated as a known anchor point that may be utilized for multi-path augmentation in pure LoS scenarios for the improved localization accuracy \cite{RIS-LOC-2, RIS-LOC-12, RIS-LOC-13}, while also functioning excellently in NLoS scenarios in the presence of obstructions by introducing beneficial LoS paths \cite{RIS-LOC-11, RIS-LOC-10}. Therefore, as can be succinctly concluded from the above attempts, RIS-aided localization is primarily characterized by the various advantages provided by the varied roles that RIS plays: i) as a known anchor point, a RIS has the potential to introduce additional location reference, based upon which the established priori facilitates the retrieval of the exact posterior associated with user's location using the Bayesian framework \cite{RIS-LOC-13,RIS-LOC-103}, and ii) due to its ability to integrate into the wireless environment functioning as any scatterer or reflector, a RIS is capable of delivering additional observations, independent of the uncontrolled and passive multiple paths, which relaxes the dependence of scatterers that is required for the standard multi-path-based positioning \cite{RIS-LOC-5}.


In spite of the potential benefits of integrating RISs in vehicular networks, achieving the high-accuracy and real-time localization remains an open issue, as evidenced by the following facts. Firstly, since the LoS channels between the vehicle and different infrastructures (e.g., BSs and RISs) originate from the same vehicle location, the LoS channel energies associated with the BS and RIS are all concentrated on the same location index, which induces a group sparsity structure of the LoS channels. The burst sparsity structure inherent to the angular NLoS channels also provides additional angle references contributing to the improved localization accuracy. Therefore, an appropriate  probabilistic model is required for characterizing the heterogeneous LoS/NLoS channels with diverse sparsities in support of the high-precision localization. Secondly, concerning that vehicle platoons are typically featured by vehicles advancing shoulder-to-shoulder at approximately the same speed, vehicle positions are highly spatial-correlated. With the passage of time, the previously obtained vehicle position delivers rich priors for the present tracking, and hence the use of such probability-dominated random correlations has the potential to achieve a high-accuracy location tracking with a reduced overhead. Furthermore, this pair of critical facts aforementioned motivate us to use sparse Bayesian-like approaches for the real-time location tracking. Nevertheless, the conventional Bayesian-based methods, e.g., variational Bayesian inference \cite{SPM-2008} and pure approximate message passing (AMP) \cite{AMP}, may be fruitless to cope with the elusive priors due to the tight coupling of the direct and cascaded location supports with various structured sparsities. Accordingly, the entries in the BS- and RIS-associated sensing matrices are not perfectly orthogonal, which renders them ill-conditioned. Consequently, a location tracking scheme has to be tailored for encapsulating LoS/NLoS channels with various structured sparsities, sufficiently robust in the presence of the sophisticated priors and highly tractable for enabling efficient algorithm design. In a nutshell, we target these non-trivial challenges in this paper to conceive a robust and high-accuracy tracking strategy for the RIS-aided vehicle platoon system.

Geared towards the challenges mentioned above, the current work presents the following advancements in the state-of-the-art:

\begin{itemize}
\item We investigate a scenario of vehicle platoon with a BS and a RIS functioning in tandem for the real-time and high-accuracy localization. The employment of a RIS imposes additional structured sparsity, which, when paired with the initial sparse LoS channels of the BS, facilitates the group sparsity. The group sparsity contributed by both enriches the energies of the original direct-only channel, resulting in a greater concentration of the LoS channel energies associated with the BS and RIS at the same location index that is based on a set of uniformed road grids prescribed by the off-grid basis. By contrast, the NLoS channels can be represented as their sparse copies, revealing the burst sparsity. This thus explicates the philosophy of the diverse sparse sparsities, which allows the original location tracking problem to be pruned as a joint estimate of the off-grid location and angle offsets. Additionally, by leveraging the geometric dilution of precision (GDOP) metric, we evince that the application of RISs can reduce the estimation error to some extent.

\item A hierarchical framework, namely \textbf{D}iverse dynam\textbf{i}c \textbf{L}ayer str\textbf{u}ctured \textbf{S}parsity (DiLuS), is crafted to concurrently capture diverse sparsities while taking into account the spatial-temporal correlations among the vehicles. 
We use a pair of layered probabilistic models to characterize these diverse sparsities inherent to the LoS/NLoS channels, each of which functions independently before being assimilated into the conditional probability of the received localization signal, thus facilitating high-accuracy location tracking using the adequate priors available. To this end, we aim at obtaining the maximum likelihood estimation (MLE) of the off-grid location and AoA offsets as well as the minimum mean square error (MMSE) estimates of the sparse LoS/NLoS channel vectors.

\item 

The investigated problem is a quasi-compressive sensing (CS) copy, but it differs significantly from the standard CS-based problem due to its ill-conditioned sensing matrices (containing the offsets to be estimated) and dynamic sparse priors furnished by DiLuS, demonstrating a high level of intractability. To attain a stable solution for this problem, we develop a DiLuS-STPL algorithm that blends the techniques of variational Bayesian inference (VBI) and message passing to recursively achieve the offset updates in a turbo-like way while taking into account the spatial-temporal correlations of the vehicle platoon. The proposed DiLuS-STPL offers the capability of fully harnessing the diverse dynamic sparse priors as enveloped by DiLuS despite the presence of uncertain sensing matrices.

\item Simulation results reveal that the RIS is a powerful facilitator for robust and high-accuracy localization by  contributing to beneficial location references, in particular for its nature of low cost and ease of implementation. We elucidate that, under the same configurations, the cooperation between a BS and a RIS achieves the comparable localization precision obtained by the two individual BSs as measured by GDOP. In addition, our proposed DiLuS-STPL algorithm performs desirably even at a coarse grid resolution, and we show that it can mitigate the NLoS-induced misleading effect while attempting to identify the LoS channels in the presence of a fluctuating number of NLoS paths, thus unveiling the critical robustness for high-accuracy localization.
\end{itemize}

The remainder of this paper is structured as follows. Section~\ref{sec-System-Model} introduces the system model and off-grid basis representation. In Section~\ref{sec-DiLuS}, we present the DiLuS framework for capturing diverse sparse priors. In Section~\ref{sec-DiLuS-STPL-Algorithm}, the DiLuS-STPL algorithm is proposed to effectively resolve the formulated location tracking problem, followed by the simulation results provided in Section~\ref{sec-Simulation}. Finally, concluding remarks are drawn in Section~\ref{sec-conclusion}.

\textit{Notation:} Scalar variables, column vectors and matrices are represented by Italic letters, boldface lower-case and capital letters such as $x$, $\bf{x}$, and $\bf{X}$, respectively. The upper-case calligraphic letters such as $\mathcal{M}$ is employed to indicate finite and discrete sets. $\mathbb{C} ^{M \times N}$ represents the space of a $ M \times N $ complex-valued matrix. 
${\left( \cdot \right)}^{*} $, ${\left( \cdot \right)}^{T} $, and ${\left( \cdot \right)}^{H} $ represent the conjugate, transpose, and Hermitian (conjugate transpose)  operators of their arguments, respectively.  
$ {\left\| {\cdot} \right\|_2} $ and $ {\left\| {\cdot} \right\|_F} $ represent the spectrum norm and Frobenius norm of their arguments, respectively. For a vector $\bf{x}$, $\text{diag}(\bf{x})$ returns a diagonal matrix with the entries of $\bf{x}$ located on its main diagonal. The symbol  $ \otimes $ stands for the Kronecker product. The imaginary unit is given by  $\jmath = \sqrt{-1} $. The expectation operator is denoted by $ \mathbb{E}\left\{  \cdot  \right\} $. Regarding $ {\cal C}{\cal N}\left( {\bf{x};\bm{\mu},\bf{\Sigma} } \right) $, we indicate that the complex random vector $\bf{x}$ is distributed in accordance with a Gaussian probability density function (PDF) with a mean vector $\bm{\mu}$ and  a covariance matrix $\bf{\Sigma}$. The symbol $  \sim  $ stands for ``distributed as".

\section{System Model}\label{sec-System-Model}
\subsection{Localization Scenario}
We consider a vehicle platoon composed of $M$ single-antenna  vehicle user equipments (VUEs) indexed by $ \mathcal{M} = \left\{ {1,...,M} \right\} $, with these VUEs moving at approximately the same velocity. The vehicle platoon is within the communication coverage of a BS equipped with a uniform linear array (ULA) having $K$ antennas, in which a RIS that is an $N$-element uniform planar array (UPA) takes over the role for enhancing the communication and localization services between the BS and VUEs. The system being investigated divides the continuous time into a number of slots indexed by $\mathcal{T} = \left\{ {1,...,T} \right\}$.  For a two-dimensional (2D) geographic area, the $m$th VUE is located at $ {\mathbf{p}}_m^t = {\left[ {x_m^t,y_m^t,z_m^t} \right]^T} $ along a road in the $t$th slot, and the center of the array gravity for the BS and the RIS are given by $ {{\mathbf{\tilde p}}^{{\text{BS}}}} = {\left[ {{x^{{\text{BS}}}},{y^{{\text{BS}}}},{z^{{\text{BS}}}}} \right]^T} $ and $ {{\mathbf{\tilde p}}^{{\text{RIS}}}} = {\left[ {{x^{{\text{RIS}}}},{y^{{\text{RIS}}}},{z^{{\text{RIS}}}}} \right]^T} $, respectively. Note that such a 2D area must be known as a priori such that the road map can be uniformed as grids to establish the exact location of each VUE. Due to the intricacy of the traffic situation, a road map suffices for the precise localization instead of the comprehensive information on the road's dynamics.


\subsection{Channel Model}

The general clustered delay line (CDL) model \cite{chen-twc2} is employed for the corresponding channels in our considered system{\footnote{ We consider a narrowband system in this paper, and the proposed algorithm can be readily extended to the wideband system. More precisely, the DiLuS paradigm still holds for the channels over each sub-carrier, and our proposed algorithm can be directly harnessed to track VUE's locations employing the measurements received from the BS at this specific sub-carrier. Furthermore, the algorithm can be extended to incorporate the ToA information in wideband system by introducing a 2D grid of both AoAs and ToAs, based upon which the ToA offsets can be likewise attained by MLE. One may refer to \cite{ICC-AnLiu} for related details in wideband systems. However, in the mobility scenario, the increased grid dimension induced by ToA information significantly enlarges the search space and exacerbates the computational burden in a short coherent interval. To facilitate simplicity and clarity, we concentrate on a narrowband system in this paper to strive for a critical trade-off between the localization performance and complexity for the RIS-aided vehicle platoon system.}}. With both highly elevated BS and RIS having a large number of antennas and elements, the spatial resolution increases significantly in the presence of the angular basis, and the channel becomes much more sparse in this case. The channels spanning from the $m$th VUE to the BS  $ {\mathbf{h}}_{m,b}^t \in {\mathbb{C}^{K \times 1}} $ and that spanning from the $m$th VUE to the RIS $ {\mathbf{h}}_{m,r}^t \in {\mathbb{C}^{N \times 1}} $ are, respectively, given by
\begin{equation}\label{channel-1}
	{\mathbf{h}}_{m,b}^t = \beta _m^t{{\mathbf{a}}_K}\left( {\omega _m^{{\text{AoA}}}} \right) + \sum\limits_{l = 1}^{{L_{m,b}}} {\tilde{\beta} _{m,l}^t{{\mathbf{a}}_K}\left( {\omega _{m,l}^{{\text{AoA}}}} \right)} ,\forall m,t,
\end{equation}
\begin{equation}\label{channel-2}
	\resizebox{1.0\hsize}{!}{$
	{\mathbf{h}}_{m,r}^t = \eta _m^t{{\mathbf{a}}_N}\left( {\varphi _m^{{\text{AoA}}},\vartheta _m^{{\text{AoA}}}} \right) + \sum\limits_{l = 1}^{{L_{m,r}}} {\tilde{\eta} _{m,l}^t{{\mathbf{a}}_N}\left( {\varphi _{m,l}^{{\text{AoA}}},\vartheta _{m,l}^{{\text{AoA}}}} \right)} ,\forall m,t.$}
\end{equation}
In (\ref{channel-1}) and (\ref{channel-2}), $ {L_{m,b}} $ and $ {L_{m,r}} $ represent the number of propagation paths for their individual channels, respectively. For the channel spanning from the $m$th VUE to the BS, $ \beta _m^t $ and $ \omega _m^{{\text{AoA}}} $ denote the complex gain and AoA of the LoS path, while $ \tilde{\beta} _{m,l}^t $ and $\omega _{m,l}^{{\text{AoA}}}$ indicate the counterparts of the $l$th NLoS path, respectively. Similarly, as for the channel spanning from the $m$th VUE to the RIS, $ \eta _m^t $ denotes the complex gain and $ \left( {\varphi _m^{{\text{AoA}}},\vartheta _m^{{\text{AoA}}}} \right) $ denotes the azimuth-elevation AoA pair, whereas $ \tilde{\eta} _{m,l}^t $ and $ \left( {\varphi _{m,l}^{{\text{AoA}}},\vartheta _{m,l}^{{\text{AoA}}}} \right)$ represents the corresponding parameters of the $l$th NLoS path. The channel complex gains associated with the LoS channels are generated according to $ {\eta _m} = \eta_0 d_{m,r} \exp \left( {\jmath \frac{{2\pi }}{\lambda }d_{m,r}} \right) $ and $ {\beta _m} = \beta_0 d_{m,b}\exp \left( {\jmath \frac{{2\pi }}{\lambda } d_{m,b} } \right) $, respectively, where $d_{m,r} = {\left\| {{\mathbf{p}}_m^t - {{{\mathbf{\tilde p}}}^{{\text{RIS}}}}} \right\|^{ - \zeta_{m,r}}}$ and $d_{m,b} = {\left\| {{\mathbf{p}}_m^t - {{{\mathbf{\tilde p}}}^{{\text{BS}}}}} \right\|^{ - \zeta_{m,b}}}$ denote their individual link distances, $ \eta_0 $ and $\beta_0$ indicate the path-loss at the reference distance of one meter, as well as  $\zeta_{m,r}$ and $\zeta_{m,b}$ for path-loss exponents. 
The NLoS counterparts are generated in compliance with Gaussian distributions with zero means and variances $ { \sigma_{\tilde{\eta}}} $ and $ { \sigma_{\tilde{\beta}}} $, i.e., $ \tilde \eta _m^t \sim \mathcal{C}\mathcal{N}\left( {0, { \sigma_{\tilde{\eta}}}  } \right) $ and $ \tilde \beta _m^t \sim \mathcal{C}\mathcal{N}\left( {0,{ \sigma_{\tilde{\beta}}} } \right) $, respectively \cite{pathloss}. The array steering vectors of the half-wavelength-space ULA and UPA at the BS and RIS are denoted by $ {{\mathbf{a}}_K}\left( \omega  \right) = {\left[ {1,{e^{ - \jmath \pi \cos \omega }},...,{e^{ - \jmath \pi \left( {K - 1} \right)\cos \omega }}} \right]^T} $ and $ {{\mathbf{a}}_N}\left( {\varphi ,\vartheta } \right) = {{\mathbf{a}}_{{N_x}}}\left( \varphi  \right) \otimes {{\mathbf{a}}_{{N_y}}}\left( \vartheta  \right) $, respectively. Furthermore, we denote $ {\mathbf{H}}_{r,b} \in {\mathbb{C}^{K \times N}} $ by the channel spanning from the RIS to the BS which is available in advance due to their fixed positions. 
The mapping between the Cartesian coordinate and the spherical coordinate regulates the conversion between corresponding angles and coordinates, e.g., $ \omega _m^{{\text{AoA}}}\left( {{\mathbf{p}}_m^t} \right) $, $ \varphi _m^{{\text{AoA}}}\left( {{\mathbf{p}}_m^t} \right)$, and  $\vartheta _m^{{\text{AoA}}}\left( {{\mathbf{p}}_m^t} \right) $, which is omitted here for conciseness. Additionally, as for the narrowband system of interest in this paper, the location information of each VUE can still be gathered from the received signal at the BS, even if no information can be retrieved from the ToA. In this case, we just need to acquire the angle-related information to determine the exact location of each VUE by applying the trick of triangulation for such a non-synchronous system~\cite{RIS-LOC-5}.

At each time slot $t$, all the VUEs simultaneously transmit $G$ information-carrying pilot sequences $ x_m^t \left( g \right), g=1,...,G, \forall t, $ for localization, in which $ x_m^t \left( g \right) $ represents the $g$th pilot sequence transmitted from the $m$th VUE in the $t$th slot. Thus, the signals propagating across the multi-path environment and being reflected by the RIS and eventually being received at the BS, are given by
\begin{align}\label{ori-signal}
{{\mathbf{y}}^t}\left( g \right) = \sum\limits_{m = 1}^M {\left( {{{\mathbf{H}}_{r,b}}{\mathbf{\Theta h}}_{m,r}^t + {\mathbf{h}}_{m,b}^t} \right)x_m^t\left( g \right) + {{\mathbf{n}}^t}\left( g \right)} ,\nonumber\\ g = 1,...,G,\forall t,
\end{align}
where $ {{\mathbf{n}}^t}\left( g \right) \sim \mathcal{C}\mathcal{N}\left( \mathbf{0}, {\sigma ^2{\mathbf{I}}} \right) $ is the additive white Gaussian noise (AWGN) with power $ \sigma ^2 $ at the BS. We denote $ {\mathbf{\Theta }} = {\text{diag}}\left( \bm{\theta } \right) \in {\mathbb{C}^{N \times N}} $ by the diagonal phase shift matrix at the RIS, where $ \bm{\theta } \in {\mathbb{C}^{N \times 1}} $ is the phase shift vector with its $n$th entry satisfying $ {\left| {{\theta _n}} \right|^2} = 1,1 \le n \le N ${\footnote{It is indeed difficult to determine a RIS phase profile in the absence of any prior channel information. What we try best is to fix it when doing channel estimation in accordance with practicable principles (e.g., random/directional RIS profiles presented in \cite{reviewer-2}).}}. Additionally, it is assumed that the large-scale parameters, such as $\eta_m^t$, $\beta_{m}^t$, $\varphi_m^{\text{AoA}}$, $\vartheta_m^{\text{AoA}}$, and $\omega_m^{\text{AoA}}$, remain approximately constant within a slot of interest. This assumption is practically reasonable since in comparison to the nominal distance with a remote BS, the traveling distance of the vehicle within one slot is marginal, resulting in  subtle variations in the geometry-associated parameters that impact the large-scale characteristics \cite{chen-jsac,chen-twc2}. This potentially relaxes the frequency of instantaneous channel information updates, thus achieving pilot savings. As regards the channel fluctuation on the slot-level of the order of hundreds of milliseconds, that of the VUE location, for instance, is on the scale of say 72~km/h (2~cm/ms).


\subsection{Off-grid Basis Representation}
Generally, the maximum likelihood (ML) or least square (LS) method may be used to directly recover the location of VUE from the received signal at the BS. However, such a strategy is arduous due to its non-convexity-induced high intractability  and the possibility of falling into a number of local optima. To resolve this issue, the location of VUE can be determined via its maximum a posteriori (MAP) estimation by exploiting the sparsity inherent to the grid basis model \cite{LOC-1}. Specifically, we introduce a uniform grid of $U$ positions along the center of the road, denoted by $\mathcal{U}$, for the location of VUE, which is given by  $ \mathcal{U} = \left\{ {{{\mathbf{r}}_1},...,{{\mathbf{r}}_U}} \right\} $. Meanwhile, a uniform grid of $ \tilde K $  AoAs at the BS over $\left[ { - \pi /2,\pi /2} \right)$  are prescribed to take values from the discrete sets: \resizebox{1.0\hsize}{!}{$ \left\{ {\tilde \omega _{m,k}^{{\text{AoA}}}:\sin \left( {\tilde \omega _{m,k}^{{\text{AoA}}}} \right) = \frac{2}{{\tilde K}}\left( {k - \left\lfloor {\frac{{\tilde K - 1}}{2}} \right\rfloor } \right),k = 0,...,\tilde K - 1,\forall m} \right\} $}. Nevertheless, the true VUE location and AoAs do not coincide exactly with the grid points. To capture the attributes of mismatches between the true angles and the grid points, we adopt an off-grid basis for the associated sparse representation. More explicitly, the AoA offset vector at the BS for the $m$th VUE is defined as $ \Delta {\bm{\omega }}_m^{{\text{AoA}}} = {\left[ {\Delta \omega _{m,1}^{{\text{AoA}}},...,\Delta \omega _{m,\tilde K}^{{\text{AoA}}}} \right]^T},\forall m $ with $ \Delta \omega _{m,k}^{{\text{AoA}}} $ given by
\begin{equation}
\Delta \omega _{m,k}^{{\text{AoA}}} = \left\{ \begin{array}{l}
	\omega _{m,k}^{{\text{AoA}}} - \tilde \omega _{m,{k_l}}^{{\text{AoA}}},k = {k_l},l = 1,...,{L_{m,b}},\forall m,\\
	0,\qquad \qquad \quad {\text{otherwise}},
\end{array} \right.
\end{equation}
where $ {k_l} $ indicates the index of the AoA grid point nearest to the AoA of the $l$th NLoS path between the  $m$th VUE and the BS. Then, let $ \Delta {\bf{r}}_m^t = \left[ {\Delta {\bf{r}}_{m,1}^t;...;\Delta {\bf{r}}_{m,U}^t} \right],\forall m $ denote the location offset vector with $ \Delta {\bf{r}}_{m,u}^t $ given by 
\begin{equation}
\Delta {\bf{r}}_{m,u}^t = \left\{ \begin{array}{l}
	{\bf{p}}_m^t - {\bf{r}}_{m,q}^t,u = q,\forall m,\\
	{\left[ {0,0} \right]^T},\quad \ \ {\text{otherwise}},
\end{array} \right.
\end{equation}
where $ q $  indicates the index of the off-grid location nearest to the $ u $th counterpart for the $m$th VUE, i.e., the square with the center of the  $ q $th off-grid location  $ {\mathbf{r}}_{m,q}^t $ in $\mathcal{U}$. Furthermore, the azimuth-elevation AoA pair $ \left( {{\bm{\varphi}} _m^{{\text{AoA}}},{\bm{\vartheta}} _m^{{\text{AoA}}}} \right) $ associated with the $m$th VUE at the RIS takes the values from a discrete set of size $ \tilde{N} $: $ \left\{ {\left( {\varphi _{m,n}^{{\text{AoA}}},\vartheta _{m,n}^{{\text{AoA}}}} \right):n = 1,...,\tilde N} \right\} $. Similarly, let $ \Delta \bm{\varphi }_m^{{\text{AoA}}} = {\left[ {\Delta \varphi _{m,1}^{{\text{AoA}}},...,\Delta \varphi _{m,\tilde N}^{{\text{AoA}}}} \right]^T},\forall m $ and $ \Delta \bm{\vartheta} _m^{{\text{AoA}}} = {\left[ {\Delta \vartheta _{m,1}^{{\text{AoA}}},...,\Delta \vartheta _{m,\tilde N}^{{\text{AoA}}}} \right]^T},\forall m $ denote the offset vectors of the azimuth-elevation AoA pair at the RIS, respectively.

With the above definitions of off-grid location and AoA offsets, the off-grid basis for the LoS/NLoS array response at the BS can be reformulated as 
\begin{align}
 {\mathbf{A}}_K^{{\text{LoS}}}\left( {\Delta {\mathbf{r}}_m^t} \right) &=  \left[ { {{\mathbf{a}}_K}\left( {\tilde \omega _{m,1}^{{\text{AoA}}}\left( {{\mathbf{r}}_{m,1}^t + \Delta {\mathbf{r}}_{m,1}^t} \right)} \right), ...,}\right. \nonumber\\ &\left. {{{\mathbf{a}}_K}\left( {\tilde \omega _{m,U}^{{\text{AoA}}}\left( {{\mathbf{r}}_{m,U}^t + \Delta {\mathbf{r}}_{m,U}^t} \right)} \right)} \right] \in \mathbb{C}^{K \times U} ,
\end{align}
\begin{align}
  {\mathbf{A}}_K^{{\text{NLoS}}}\left( {\Delta \bm{\omega }_{m}^{{\text{AoA}}}} \right) &= \left[ {{{\mathbf{a}}_K}\left( {\tilde \omega _{m,1}^{{\text{AoA}}} + \Delta \omega _{m,1}^{{\text{AoA}}}} \right), ..., }\right.  \nonumber \\  &\left. {{{\mathbf{a}}_K}\left( {\tilde \omega _{m,\tilde K}^{{\text{AoA}}} + \Delta \omega _{m,\tilde K}^{{\text{AoA}}}} \right)} \right] \in {\mathbb{C}^{K \times \tilde K}}.
\end{align}
Similarly, the off-grid basis for the LoS/NLoS array response at the RIS can be given by
\begin{subequations}
\begin{align}
&{\mathbf{A}}_N^{{\text{LoS}}}\left( {\Delta {\mathbf{r}}_m^t} \right) \nonumber\\& = \left[ {{{\mathbf{a}}_N}\left( {\varphi _{m,1}^{{\text{AoA}}}\left( {{\mathbf{r}}_{m,1}^t + \Delta {\mathbf{r}}_{m,1}^t} \right),\vartheta _{m,1}^{{\text{AoA}}}\left( {{\mathbf{r}}_{m,1}^t + \Delta {\mathbf{r}}_{m,1}^t} \right)} \right), }\right. \nonumber \\  
&\quad  
\left. {...,{{\mathbf{a}}_N}\left( {\varphi _{m,U}^{{\text{AoA}}}\left( {{\mathbf{r}}_{m,U}^t + \Delta {\mathbf{r}}_{m,U}^t} \right),\vartheta _{m,U}^{{\text{AoA}}}\left( {{\mathbf{r}}_{m,U}^t + \Delta {\mathbf{r}}_{m,U}^t} \right)} \right)} \right] \nonumber\\ &  \in {\mathbb{C}^{N \times U}}  ,\\
 &{\mathbf{A}}_N^{{\text{NLoS}}}\left( {\Delta \bm{\varphi }_m^{{\text{AoA}}},\Delta \bm{\vartheta} _m^{{\text{AoA}}}} \right) \nonumber\\ 
 & = \left[ {{{\mathbf{a}}_N}\left( {\tilde \varphi _{m,1}^{{\text{AoA}}} + \Delta \varphi _{m,1}^{{\text{AoA}}},\tilde \vartheta _{m,1}^{{\text{AoA}}} + \Delta \vartheta _{m,1}^{{\text{AoA}}}} \right), }\right. \nonumber \\ 
 & \quad  
  \left. { ...,{{\mathbf{a}}_N}\left( {\tilde \varphi _{m,\tilde N}^{{\text{AoA}}} + \Delta \varphi _{m,\tilde N}^{{\text{AoA}}},\tilde \vartheta _{m,\tilde N}^{{\text{AoA}}} + \Delta \vartheta _{m,\tilde N}^{{\text{AoA}}}} \right)} \right] \in {\mathbb{C}^{N \times \tilde N}}  .
\end{align}
\end{subequations}
Therefore, the signal received from $M$ VUEs at the BS can be recast as a CS form with indeterminate parameters in the sensing matrix, i.e.,
\begin{equation}\label{CS}
{{\mathbf{y}}^t} = \left[ {\begin{array}{*{20}{c}}
		{{{\mathbf{F}}^t}\left( {\Delta {{\mathbf{r}}^t}} \right)}&{{{\mathbf{\Xi }}^t}\left( {\Delta {\bm{\omega }^{{\text{AoA}}}},\Delta {\bm{\varphi }^{{\text{AoA}}}},\Delta {\bm{\vartheta} ^{{\text{AoA}}}}} \right)} 
\end{array}} \right]\left[ {\begin{array}{*{20}{c}}
		{{{\mathbf{z}}^t}} \\ 
		{{{\mathbf{v}}^t}} 
\end{array}} \right] + {{\mathbf{n}}^t},
\end{equation}
where $ {{\mathbf{y}}^t} = \left[ {{{\mathbf{y}}^t}\left( 1 \right);...;{{\mathbf{y}}^t}\left( G \right)} \right]  \in {\mathbb{C}^{KG \times 1}}  $, $ {{\mathbf{n}}^t} = \left[ {{{\mathbf{n}}^t}\left( 1 \right);...;{{\mathbf{n}}^t}\left( G \right)} \right] \in {\mathbb{C}^{KG \times 1}} $, $ {{\mathbf{z}}^t} = \left[ {{\mathbf{z}}_{{\text{R}},1}^t;...;{\mathbf{z}}_{{\text{R}},M}^t; }\right. \left. { {\mathbf{z}}_{{\text{B}},1}^t;...;{\mathbf{z}}_{{\text{B}},M}^t} \right] \in {\mathbb{C}^{ 2UM  \times 1}} $, and $ {{\mathbf{v}}^t} = \left[ {{\mathbf{v}}_{{\text{R}},1}^t;...;{\mathbf{v}}_{{\text{R}},M}^t;{\mathbf{v}}_{{\text{B}},1}^t;...;{\mathbf{v}}_{{\text{B}},M}^t} \right] \in {\mathbb{C}^{\left( {\tilde K + \tilde N} \right)M \times 1}} $. The sensing matrices $ {{{\mathbf{F}}^t}\left( {\Delta {{\mathbf{r}}^t}} \right)} $ and ${{{\mathbf{\Xi }}^t}\left( {\Delta {\bm{\omega }^{{\text{AoA}}}},\Delta {\bm{\varphi }^{{\text{AoA}}}},\Delta {{\bm{\vartheta}} ^{{\text{AoA}}}}} \right)}$ are expressed as
\begin{equation}
 {{\mathbf{F}}^t}\left( {\Delta {{\mathbf{r}}^t}} \right) = \left[ {\begin{array}{*{20}{c}}
		{{{\mathbf{H}}_{r,b}}{\mathbf{\Theta A}}_N^{{\text{LoS}}}\left( 1 \right)}&{{\mathbf{A}}_K^{{\text{LoS}}}\left( 1 \right)} \\ 
		\vdots & \vdots  \\ 
		{{{\mathbf{H}}_{r,b}}{\mathbf{\Theta A}}_N^{{\text{LoS}}}\left( G \right)}&{{\mathbf{A}}_K^{{\text{LoS}}}\left( G \right)} 
\end{array}} \right] \in {\mathbb{C}^{KG \times 2UM}} , 
\end{equation}
\begin{align}
&{{\mathbf{\Xi }}^t}\left( {\Delta {\bm{\omega }^{{\text{AoA}}}},\Delta {\bm{\varphi }^{{\text{AoA}}}},\Delta {{\bm{\vartheta}} ^{{\text{AoA}}}}} \right)  \nonumber\\ &= \left[ {\begin{array}{*{20}{c}}
		{{{\mathbf{H}}_{r,b}}{\mathbf{\Theta A}}_N^{{\text{NLoS}}}\left( 1 \right)}&{{\mathbf{A}}_K^{{\text{NLoS}}}\left( 1 \right)} \\ 
		\vdots & \vdots  \\ 
		{{{\mathbf{H}}_{r,b}}{\mathbf{\Theta A}}_N^{{\text{NLoS}}}\left( G \right)}&{{\mathbf{A}}_K^{{\text{NLoS}}}\left( G \right)} 
\end{array}} \right] \in {\mathbb{C}^{KG \times \left( {\tilde K + \tilde N} \right)M}},
\end{align}
where the entries $ {{\mathbf{H}}_{r,b}}{\mathbf{\Theta A}}_N^{{\text{LoS}}}\left( g \right) $, $ {\mathbf{A}}_K^{{\text{LoS}}}\left( g \right) $, $ 	{{\mathbf{H}}_{r,b}}{\mathbf{\Theta A}}_N^{{\text{NLoS}}}\left( g \right) $, and $ {\mathbf{A}}_K^{{\text{NLoS}}}\left( g \right) $ are shown in (\ref{tmp-1}) at the top of the next page.
\begin{figure*}[!htb]
\begin{subequations}\label{tmp-1}
\begin{align}
{{\mathbf{H}}_{r,b}}{\mathbf{\Theta A}}_N^{{\text{LoS}}}\left( g \right) & = {\left[ {x_1^t\left( g \right){{\mathbf{H}}_{r,b}}{\mathbf{\Theta A}}_N^{{\text{LoS}}}\left( {\Delta {{\mathbf{r}}^t}} \right),...,x_M^t\left( g \right){{\mathbf{H}}_{r,b}}{\mathbf{\Theta A}}_N^{{\text{LoS}}}\left( {\Delta {{\mathbf{r}}^t}} \right)} \right] \in \mathbb{C}^ {K \times MU}}, \\
{\mathbf{A}}_K^{{\text{LoS}}}\left( g \right) &=  {\left[ {x_1^t\left( g \right){\mathbf{A}}_K^{{\text{LoS}}}\left( {\Delta {{\mathbf{r}}^t}} \right),...,x_M^t\left( g \right){\mathbf{A}}_K^{{\text{LoS}}}\left( {\Delta {{\mathbf{r}}^t}} \right)} \right] \in \mathbb{C}^ {K \times MU}}, \\
	{{\mathbf{H}}_{r,b}}{\mathbf{\Theta A}}_N^{{\text{NLoS}}}\left( g \right)  &= \left[ {x_1^t\left( g \right){{\mathbf{H}}_{r,b}}{\mathbf{\Theta A}}_N^{{\text{NLoS}}}\left( {\Delta \bm{\varphi }_1^{{\text{AoA}}},\Delta {\bm{\vartheta}} _1^{{\text{AoA}}}} \right), } \right. \nonumber \\ 
	& \qquad \qquad \qquad  
	\left. {   ...,x_M^t\left( g \right){{\mathbf{H}}_{r,b}}{\mathbf{\Theta A}}_N^{{\text{NLoS}}}\left( {\Delta \bm{\varphi }_M^{{\text{AoA}}},\Delta {\bm{\vartheta}} _M^{{\text{AoA}}}} \right)} \right] \in \mathbb{C} ^{K \times M\tilde N} ,\\
{\mathbf{A}}_K^{{\text{NLoS}}}\left( g \right) &= \left[ {x_1^t\left( g \right){\mathbf{A}}_K^{{\text{NLoS}}}\left( {\Delta \bm{\omega }_1^{{\text{AoA}}}} \right),  ..., } \right.  \left.{ x_M^t\left( g \right){\mathbf{A}}_K^{{\text{NLoS}}}\left( {\Delta \bm{\omega }_{K \times \tilde K}^{{\text{AoA}}}} \right)} \right] \in \mathbb{C}^{K \times M\tilde K}.
\end{align}
\end{subequations}
\hrule
\end{figure*}
Next, let us concentrate on the physical meaning of the two blocks, i.e., ${\bf{z}}^t$ and ${\bf{v}}^t$, in the sparse vector $ {\left[ {\begin{array}{*{20}{c}}
			{{{\mathbf{z}}^t}}&{{{\mathbf{v}}^t}} 
\end{array}} \right]^T} $. As for ${\bf{z}}^t$, the vector $ {\mathbf{z}}_{{\text{R}},m}^t = \left[ z_{{\text{R}},m,1}^t,...,z_{{\text{R}},m,U}^t \right] \in {\mathbb{C}^{U \times 1}} $ represents the sparse LoS channel pertaining to the RIS-VUE link, the $u$th entry of which denote the complex gain of the LoS path associated with the off-grid location ${\mathbf{r}}_{m,u}^t + \Delta {\mathbf{r}}_{m,u}^t$ in the $t$th slot. The vector $ {\mathbf{z}}_{{\text{B}},m}^t \in \mathbb{C}^{U \times 1} $ represents the sparse LoS channel for the BS-VUE link. Note that each ${\mathbf{z}}_{{\text{R}},m}^t$ and ${\mathbf{z}}_{{\text{B}},m}^t$ only has one non-zero entry that points to the true location of the $m$th VUE, whose index of the non-zero entries of LoS channel vectors $\left[ {{\mathbf{z}}_{{\text{R}},m}^t;{\mathbf{z}}_{{\text{B}},m}^t} \right], \forall m$, are the same, as given by $q_m^t$ (which will be clear later). In other words, the LoS channel vectors $\left[ {{\mathbf{z}}_{{\text{R}},m}^t;{\mathbf{z}}_{{\text{B}},m}^t} \right], \forall m$, can be grouped into $U$ blocks with block size 2, with the $u$th block constituted by $ \left[ {z_{{\text{R}},m,u}^t,z_{{\text{B}},m,u}^t} \right], \forall m $, and only the $u$th block is non-zero. This implies that the sparse LoS channel vector $ {\bf{z}}^t $ can be grouped into $UM$ blocks with block size~2. By contrast, the sparse vectors $ {\bf{v}} _{\text{R},m} ^t $ and $ {\bf{v}} _{\text{B},m} ^t, \forall m $, contained in the block $ {\bf{v}} ^t $ denotes the complex gains of NLoS paths with respect to direct and cascaded channels, respectively.
$ {\mathbf{v}}_{{\text{R}},m}^t = \left[ {{\mathbf{v}}_{{\text{R}},m,1}^t,...,{\mathbf{v}}_{{\text{R}},m,\tilde K}^t} \right] \in {\mathbb{C}^{\tilde K \times 1}} $ stands for the NLoS channel vector, the $k$th entry of which denotes the complex gain of the NLoS path arriving at the RIS with elevation-azimuth pair $\left( {\tilde \varphi _{m,n}^{{\text{AoA}}} + \Delta \varphi _{m,n}^{{\text{AoA}}},\tilde \vartheta _{m,n}^{{\text{AoA}}} + \Delta \vartheta _{m,n}^{{\text{AoA}}}} \right)$. The NLoS channel vector $ {\mathbf{v}}_{{\text{B}},m}^t $ associated with the BS conform to the same definition given above. Note that there are $L_{m,r}$ and $L_{m,b}$ non-zero entries in ${\mathbf{v}}_{{\text{R}},m}^t $  and  $ {\mathbf{v}}_{{\text{B}},m}^t$ , respectively, in which the indices of the non-zero entries indicate the significant AoAs associated with the NLoS paths. Therefore, the diverse sparsities inherent in (\ref{CS}) motivate us to conceive a customized tracking framework for the RIS-aided vehicle platoons.

\subsection{Geometric Dilution of Precision (GDOP) Metric}
GDOP has been commonly used as a metric for the accuracy evaluation of localization and tracking systems. Due to the fact that the high-accuracy localization requires both precise range measurement and a strong geometric association between the VUE and prescribed grids, the study of GDOP is a crucial aspect of assessing the performance of a localization system. Despite the fact that the Cram{\'e}r-Rao lower bound (CRLB) metric may be applied to represent the minimum variance of the estimate error from any unbiased estimator, it fails to explicitly quantify the impact of system geometry on performance and also incorporates the effect of input noise~\cite{RIS-LOC-5-71}. Therefore, we elucidate that the integration of RIS into our considered vehicle platoon system has a beneficial GDOP performance by means of the concise GDOP analytical expressions for each VUE. We first present the GDOP in the instance of a ULA at the BS, and then extend it to the UPA case at the RIS, as these phases are pertinent to the discussion that follows.

\subsubsection{ULA Case}
Let $ {\mu _{{\text{B}},k}} $ denote the $k$th entry of ${\mathbf{h}}_{m,b}^t$ when there is no ambiguity, in which case the Fisher information matrix (FIM) associated with the AoA $\omega _m^{{\text{AoA}}}$ returns to a scalar 
\begin{equation}\label{FIM_omega}
{\text{FIM}}\left( {\omega _m^{{\text{AoA}}}} \right) = \frac{2}{{{\sigma ^2}}}\operatorname{Re} \left\{ {\sum\limits_{k = 1}^K {\frac{{\partial \mu _{{\text{B}},k}^*}}{{\omega _m^{{\text{AoA}}}}}\frac{{\partial {\mu _{{\text{B}},k}}}}{{\omega _m^{{\text{AoA}}}}}} } \right\},
\end{equation}
with $ \frac{{\partial {\mu _{{\text{B}},k}}}}{{\omega _m^{{\text{AoA}}}}} = {\eta _m}\pi \left( {k - 1} \right)\cos \omega _m^{{\text{AoA}}}\exp \left\{ { - \jmath \pi \left( {k - 1} \right)\sin \omega _m^{{\text{AoA}}} }\right.  \\ \left. { - \jmath \frac{\pi }{2}} \right\} $ and $ \frac{{\partial \mu _{{\text{B}},k}^*}}{{\omega _m^{{\text{AoA}}}}} = \eta _m^*\pi \left( {k - 1} \right)\cos \omega _m^{{\text{AoA}}} \exp  \left\{ { - \jmath \pi \left( {k - 1} \right) \times }\right.  \\ \left. { \sin \omega _m^{{\text{AoA}}} + \jmath \frac{\pi }{2}} \right\} $. We can further structure (\ref{FIM_omega}) in the form of
\begin{align}
	{\text{FIM}}\left( {\omega _m^{{\text{AoA}}}} \right) &= \frac{2}{{{\sigma ^2}}}{\left( {\pi \cos \omega _m^{{\text{AoA}}}} \right)^2}{\left| {{\eta _m}} \right|^2}\sum\limits_{k = 1}^K {{{\left( {k - 1} \right)}^2}} \nonumber \\
	  &\propto \frac{{{K^3}}}{3}\frac{2}{{{\sigma ^2}}}{\left( {\pi \cos \omega _m^{{\text{AoA}}}} \right)^2}{\left| {{\eta _m}} \right|^2}.
\end{align}
Thus, the CRLB associated with the AoA $\omega _m^{{\text{AoA}}}$ at the BS can be given by
\begin{align}\label{CRLB_ULA}
{\text{CRLB}}\left( {\omega _m^{{\text{AoA}}}} \right) &= {\text{FI}}{{\text{M}}^{ - 1}}\left( {\omega _m^{{\text{AoA}}}} \right) \nonumber\\&= \frac{{3{\sigma ^2}}}{{2{K^3}{{\left( {\pi \cos \omega _m^{{\text{AoA}}}} \right)}^2}{{\left| {{\eta _m}} \right|}^2}}}.
\end{align}
Furthermore, the GDOP remapped from the $\omega _m^{{\text{AoA}}}$-based CRLB at the BS can be expressed as
\begin{align}
	{\text{GDO}}{{\text{P}}_{{\text{ULA}}}} &= {\left( {\cos \omega _m^{{\text{AoA}}}} \right)^{ - 1}}d_{m,b}^{ - {\zeta _{m,b}}}\sqrt {{\text{CRLB}}\left( {\omega _m^{{\text{AoA}}}} \right)} \nonumber\\
	  &= \sqrt {1.5} \sigma {K^{ - 1.5}}\eta _0^{ - 1}{\pi ^{ - 1}}d_{m,b}^{1 + {\zeta _{m,b}}}{\left( {\cos \omega _m^{{\text{AoA}}}} \right)^{ - 1}}.
\end{align}
Without loss of generality, $ {\eta _0} = {\left( {c/4\pi {f_c}} \right)^{{\zeta _{m,b}}}} $ with $c$ being the speed of light and $f_c$ for the carrier frequency, and regarding the half-wavelength-space ULA, ${\text{GDOP}}_{\text{ULA}}$ can be recast as
\begin{align}\label{GDOP_ULA}
&{\text{GDO}}{{\text{P}}_{{\text{ULA}}}}  \nonumber\\ &= \sqrt {1.5} {\pi ^{ - 1}}{\left( {c/4\pi } \right)^{ - {\zeta _{m,b}}}}\sigma {K^{ - 1.5}}f_c^{{\zeta _{m,b}}}d_{m,b}^{1 + {\zeta _{m,b}}}{\left( {\cos \omega _m^{{\text{AoA}}}} \right)^{ - 1}} \nonumber\\
& \triangleq {\text{GDOP}}\left( {\omega _m^{{\text{AoA}}},{d_{m,b}};\sigma ,K,{f_c},{\zeta _{m,b}}} \right).
\end{align}
As observed in (\ref{GDOP_ULA}), the GDOP is dependent on different factors, e.g., AoAs at the BS $\omega_m ^{\text{AoA}}$, the link distance $d_{m,b}$, the noise $\sigma$, the array size $K$, the carrier frequency $f_c$, and the path-loss exponent $\zeta_{m,b}$, etc. Next, we extend the aforementioned insights to the instance of a UPA at the RIS.

\subsubsection{UPA Case}
For an $N$-element UPA, we further denote $N_h$ and $N_v$ by the number of elements along the horizontal and vertical directions, respectively, i.e., $N = N_h \times N_v$, in which we can readily obtain the GDOP of the $ n_v $th ULA with size $N_h$ based on the argument in (\ref{GDOP_ULA}). In view of the independence among the observations achieved by each ULA, the observation of a UPA can be treated as a combination of $N_v$ observations, thus yielding that
\begin{align}\label{GDOP_UPA}
{\text{GDO}}{{\text{P}}_{{\text{UPA}}}} = & \sqrt {1.5} {\pi ^{ - 1}}{\left( {c/4\pi } \right)^{ - {\zeta _{m,r}}}}\sigma    N_h^{ - 1.5}N_v^{ - 0.5} \nonumber\\ & \times f_c^{{\zeta _{m,r}}}d_{m,r}^{1 + {\zeta _{m,r}}}{\left( {\cos \varphi _m^{{\text{AoA}}}\cos \vartheta _m^{{\text{AoA}}}} \right)^{ - 1}} .
\end{align}

\subsubsection{GDOP Achieved by a BS Plus a RIS}
In light of the facts in (\ref{GDOP_ULA}) and (\ref{GDOP_UPA}), the cascaded channel  $ {{\mathbf{H}}_{r,b}}{\mathbf{\Theta h}}_{m,r}^t $-associated GDOP at the RIS can be given by
\begin{align}\label{GDOP_RIS}
{\text{GDO}}{{\text{P}}_{{\text{RIS}}}} =& {\left| {{\text{P}}{{\text{L}}_{r,b}}} \right|^{ - 1}}{K^{ - 0.5}}N_v^{ - 0.5} \nonumber\\  & \times {\text{GDOP}}\left( {\varphi _m^{{\text{AoA}}},\vartheta _m^{{\text{AoA}}},{d_{m,r}};\sigma ,{N_h},{f_c},{\zeta _{m,r}}} \right),
\end{align}
where we use ${\text{P}}{{\text{L}}_{r,b}}$ here for representing the complex gain of the BS-RIS channel. One can refer to Appendix~\ref{Appendix_GDOP} for detailed derivation. Accordingly, we arrive at the closed-form GDOP expression with respect to the implementation of a BS plus a RIS
\begin{equation}
	 {\text{GDO}}{{\text{P}}_{{\text{BS\&RIS}}}} = {\left( {{\text{GDOP}}_{{\text{BS}}}^{ - 1} + {\text{GDOP}}_{{\text{RIS}}}^{ - 1}} \right)^{ - 1}} .
\end{equation}
Given that ${\text{GDO}}{{\text{P}}_{{\text{RIS}}}}$ is typically non-negative, it is evident that  
\begin{equation}
 {\text{GDO}}{{\text{P}}_{{\text{BS\&RIS}}}} \le {\text{GDO}}{{\text{P}}_{{\text{BS}}}} .
\end{equation}
This implies that the application of RIS evidently improves localization precision owing to the richer localization information explored by RIS, i.e., additional LoS paths dominated by the significant angles, resulting in a reduction in error estimation (a lower GDOP).

Additionally, one may argue that the implementation of double BSs results in possibly better localization performance in comparison to the case of a BS plus a RIS. Despite its intuitive plausibility, achieving this typically necessitates a number of ideal conditions, such as the presence of strong direct links without any blockages, which may be impractical due to the complex nature of traffic patterns. Putting another BS in a single cell may introduce significant interference and additional complexity of the practical implementation since a BS is far more expensive than a RIS. From an economic standpoint, we would favor a more cost-effective solution if the two function comparably in the majority of circumstances. Then, to compare the localization performance of double BSs and a BS plus a RIS, we analyze their GDOPs in the following.

\subsubsection{A BS Plus a RIS vs. Double BSs}
The GDOP with respect to the implementation of double BSs can be explicitly written as
\begin{equation}
{\text{GDOP}} = {\left( {{\text{GDOP}}_{{\text{BS}}}^{ - 1} + {\text{GDOP}}_{{\text{BS}}}^{ - 1}} \right)^{ - 1}} = 0.5{\text{GDO}}{{\text{P}}_{{\text{BS}}}}.
\end{equation}
Given the facts in (\ref{GDOP_ULA}) and (\ref{GDOP_RIS}), $ {\text{GDO}}{{\text{P}}_{{\text{BS}}}}/{\text{GDO}}{{\text{P}}_{{\text{RIS}}}} $ can be given by (\ref{GDOP_ratio}) shown at the top of the next page.
\begin{figure*}[!htb]
\begin{equation} \label{GDOP_ratio}
	\frac{{{\text{GDO}}{{\text{P}}_{{\text{BS}}}}}}{{{\text{GDO}}{{\text{P}}_{{\text{RIS}}}}}} = \frac{{\sqrt {1.5} {\pi ^{ - 1}}{{\left( {c/4\pi } \right)}^{ - {\zeta _{m,b}}}}\sigma {K^{ - 1.5}}f_c^{{\zeta _{m,b}}}d_{m,b}^{1 + {\zeta _{m,b}}}{{\left( {\cos \omega _m^{{\text{AoA}}}} \right)}^{ - 1}}}}{{{{\left| {{\text{P}}{{\text{L}}_{r,b}}} \right|}^{ - 1}}{K^{ - 0.5}}N_v^{ - 0.5}{\text{GDOP}}\left( {\varphi _m^{{\text{AoA}}},\vartheta _m^{{\text{AoA}}},{d_{m,r}};\sigma ,{N_h},{f_c},{\zeta _{m,r}}} \right)}}.
\end{equation}
\hrule
\end{figure*}
It is intuitive that the ratio $ {\text{GDO}}{{\text{P}}_{{\text{BS}}}}/{\text{GDO}}{{\text{P}}_{{\text{RIS}}}} $ is subject to a number of parameters associated with the direct and cascaded channels, in particular for the array scale and the link distance. If we appropriately neglect some trivial parameters, (\ref{GDOP_ratio}) can be boldly trimmed to
\begin{equation}\label{trimmed_GDOP_ration}
	\frac{{{\text{GDO}}{{\text{P}}_{{\text{BS}}}}}}{{{\text{GDO}}{{\text{P}}_{{\text{RIS}}}}}} \propto \left| {{\text{P}}{{\text{L}}_{r,b}}} \right| \cdot {K^{ - 1}}N_h^{1.5}N_v^{0.5}.
\end{equation}
It can be inferred that the localization performance achieved by a BS plus a RIS is superior to that of the double BSs in the event of the ratio $ {\text{GDO}}{{\text{P}}_{{\text{BS}}}}/{\text{GDO}}{{\text{P}}_{{\text{RIS}}}} $ being greater than 1, i.e., $ \frac{{{\text{GDO}}{{\text{P}}_{{\text{BS}}}}}}{{{\text{GDO}}{{\text{P}}_{{\text{RIS}}}}}} > 1 $. With a glance at (\ref{trimmed_GDOP_ration}), the ratio $  {\text{GDO}}{{\text{P}}_{{\text{BS}}}}/{\text{GDO}}{{\text{P}}_{{\text{RIS}}}}  $ is inversely proportional to $ K $ while proportionate to $ N_h $ and  $ N_v $. This connotes that the localization performance of RIS may be considerably enhanced by increasing its element number $N = N_h \times N_v$, since this may compensate for the multiplicative fading nature inherent to the cascaded channels \cite{chen-jsac}. Therefore, it is not wise to follow empiricism rather than observing their attributes with theoretical analyses. Under different simulation configurations, we compare localization performance attained by double BS and a BS plus RIS using GDOP metric, as will be elaborated in Sec.~V.~C.

\begin{figure*}[t]
	\centering
	\includegraphics[width=\textwidth]{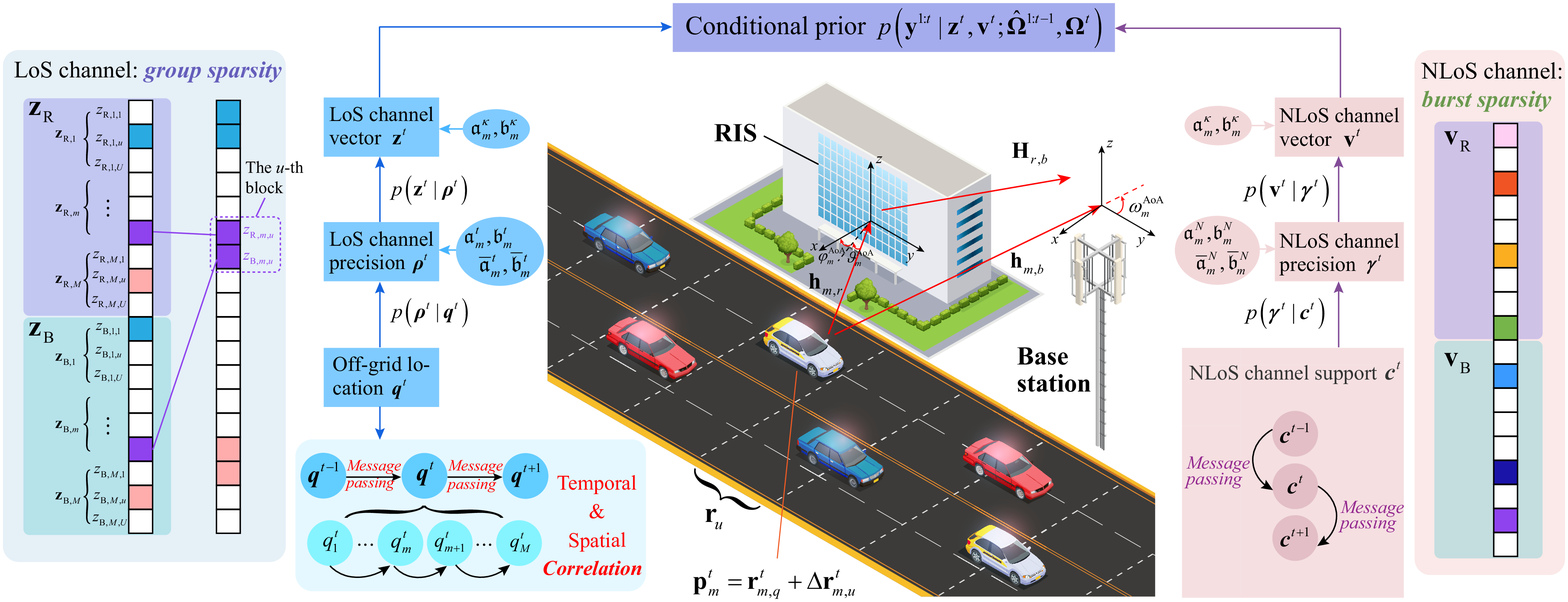}
	\caption{The proposed DiLuS framework.} \label{DiLuS}
\end{figure*}

\section{Diverse Dynamic Layered Structured Sparsity (DiLuS) Framework}\label{sec-DiLuS}
In regard to high-accuracy and robust VUE location tracking, we intend to recover both the sparse LoS and NLoS channel vectors, namely $ {\mathbf{z}}^t $ and $ {\mathbf{v}}^t $ as presented in (\ref{CS}), for each VUE of the considered vehicle platoon system from the observation $\mathbf{y}_m^t$. Although such a parameter-recovery problem seems straightforward, it poses a few challenges owing to a pair of peculiarities: i) the uncertainty of the sensing matrices containing the location and angle offsets to be determined, i.e., $ {\mathbf{F}}_m^t $ and $ {\mathbf{\Xi }}_m^t $, and ii) the unavailability of the exact distributions for the sparse vectors $ {\mathbf{z}}^t $ and $ {\mathbf{v}}^t $ with diverse sparsities (the same for the distribution of each entry within both vectors). Therefore, targeting the twofold issues, we customize a DiLuS framework as illustrated in Fig.~\ref{DiLuS}, for completely capturing diverse sparsity structures. The proposed DiLuS, on the one hand, leverages spatial-temporal correlations associated with LoS/NLoS channels among the vehicle platoon, in a formation to facilitate the closed-form updates with reduced complexity and overhead. On the other hand, it can provide appropriate prior distributions for $ {\mathbf{z}}^t $ and $ {\mathbf{v}}^t $ in support of efficient and robust performance in location tracking across a range of channel distributions.

\subsection{LoS Channel}
Recall that $ {{\mathbf{z}}^t} $ represents the sparse LoS channels incorporating the information of both direct and cascaded channels, and we let $ {\bm{\rho }^t} = \left[ {\bm{\rho }_{\text{R}}^t;\bm{\rho }_\text{B}^t} \right] \in {\mathbb{C}^{2UM \times 1}} $ represent the precision vector of $ {{\mathbf{z}}^t} $. In particular, $ \bm{\rho }_{\text{R}}^t = \left[ {\bm{\rho }_{{\text{R}},1}^t;...;\bm{\rho }_{{\text{R}},M}^t} \right] \in {\mathbb{C}^{UM \times 1}} $, $ \bm{\rho }_{\text{B}}^t = \left[ {\bm{\rho }_{{\text{B}},1}^t;...;\bm{\rho }_{{\text{B}},M}^t} \right] \in {\mathbb{C}^{UM \times 1}} $, $ \bm{\rho }_{{\text{R}},m}^t = {\left[ {\rho _{{\text{R}},m,1}^t,...,\rho _{{\text{R}},m,U}^t} \right]^T} \in {\mathbb{C}^{U \times 1}} $,  and $ \bm{\rho }_{{\text{B}},m}^t = {\left[ {\rho _{{\text{B}},m,1}^t,...,\rho _{{\text{B}},m,U}^t} \right]^T} \in {\mathbb{C}^{U \times 1}} $, where $ \rho _{{\text{R}},m,u}^t $ indicates the precision of the entry $ z _{\text{R},m,u}^t $ involved in the $u$th block with $ 1/\rho _{\text{R},m,u}^t $ being the variance of $ z _{\text{R},m,u}^t $, and $ \rho _{{\text{B}},m,u}^t $ for $ z_{{\text{B}},m,u}^t$. For ease of exposition, we denote a sequence of time-related variables $ \left\{ {{{\mathbf{z}}^\tau }} \right\}_{\tau  = 1}^t$ by $ {{\mathbf{z}}^{1:t}} $, the shorthand of which is shared with others.

\subsubsection{Dynamic Hidden Markov Model for the Vehicle Platoon}
Given that $ q $ indicates the index of the off-grid location nearest to the $u$th prescribed grid for the $m$th VUE, we refine $ q_m^t \in \left\{ {1,...,U} \right\} $ as the off-grid location state in order to represent the coarse position of the $m$th VUE along the prescribed $U$ location grids in the $t$th slot, with ${\bm{q}^t} = {\left[ {q_1^t,...,q_M^t} \right]^T},\forall t$. Taking into account the spatial-temporal attributes of vehicle platoons, a dynamic hidden Markov model is used to generate priors, i.e.,
\begin{align}
	p\left( {{{\bm{q}}^{1:t}}} \right) &= \prod\limits_\tau  {p\left( {{{\bm{q}}^\tau }|{{\bm{q}}^{\tau  - 1}}} \right)} \nonumber\\ &= \prod\limits_\tau  {\prod\limits_m {\underbrace {p\left( {q_{m + 1}^\tau |q_m^\tau } \right)}_{ \triangleq h_{S,m}^\tau \left( {{\text{Spatial}}} \right)}\underbrace {p\left( {q_m^\tau |q_m^{\tau  - 1}} \right)}_{ \triangleq h_{T,m}^\tau \left( {{\text{Temporal}}} \right)}} }.
\end{align}
Particularly, we suppose that the spatial correlation between the $m$th and $(m+1)$th VUE, i.e., $p \left( {q_{m+1}^\tau | q_m^\tau} \right) $, is drawn from a sampled Gamma distribution \cite{VP-1}
\begin{equation}\label{location-gamma}
	p\left( {q_{m + 1}^\tau |q_m^\tau } \right) \propto \frac{{{{\left( {q_m^\tau  - {q_0}} \right)}^{\varpi  - 1}}}}{{{\lambda ^\varpi }\Gamma \left( \varpi  \right)}}\exp \left( { - \frac{{q_m^\tau  - {q_0}}}{\lambda }} \right),q_m^\tau  \ge {q_0},
\end{equation}
where $ {q_0} $ denotes the minimum distance between vehicles, $ \varpi  \ge 1 $ is the common shape parameter, and $ \lambda $ is the scale parameter. Exploiting the collected data (received signals) and the EM technique, parameters $ \varpi $ and $ \lambda $ can be calculated in practice, which is a widely held belief in the academic literature confirmed by empirical research~\cite{LOC-1,VP-1}.
Furthermore, the VUE's location in the previous slot provides a priori indicator of its location in the present slot, which may be achieved by message passing and will be expounded upon later in Sec.~\ref{sec-DiLuS-STPL-Algorithm}.~B.

\subsubsection{Layers II \& III: A Pair of Conjugate Distributions}
A Gamma prior distribution is imposed on the precision vector $ {{\bm{\rho }}^t} $  of the off-grid location state $ {\bm{q}^t} $
\begin{equation}\label{layer-II}
p\left( {{\bm{\rho }^t}|{\bm{q}^t}} \right) = \prod\limits_{m,u} {p\left( {\rho _{\text{R},m,u}^t|q_m^t} \right)p\left( {\rho _{\text{B},m,u}^t|q_m^t} \right)} ,
\end{equation}
where $ p\left( {\rho _{i,m,u}^t|q_m^t} \right) = \prod\limits_{m,u} \Gamma {{\left( {\rho _{i,m,u}^t;\mathfrak{a}_{i,m}^t,\mathfrak{b}_{i,m}^t} \right)}^{\mathbb{I}\left( {q = u} \right)}} \times \\ \Gamma {{\left( {\rho _{i,m,u}^t;\bar {\mathfrak{a}}_{i,m}^t,\bar {\mathfrak{b}}_{i,m}^t} \right)}^{\mathbb{I}\left( {q \ne u} \right)}} ,i \in \left\{ {\text{R},\text{B}} \right\} $, and $ \Gamma \left( {\rho ;\mathfrak{a},\mathfrak{b}} \right) $ is a Gamma hyperprior with the shape parameter $\mathfrak{a}$ and rate parameter $ \mathfrak{b} $. For the $m$th VUE, $ {\mathfrak{a}_{i,m}} $ and  $ {\mathfrak{b}_{i,m}} $ are the shape and rate parameters of the LoS channel precision $ \rho _{i,m,u}^t $, conditioned on the fact that the  $ u $th off-grid location corresponds to the $m$th VUE’s location, i.e.,  $ q = u $. On the contrary, the parameter and rate parameters  $ {\bar{ \mathfrak{a}}_{i,m}} $  and  $ {\bar{ \mathfrak{b}}_{i,m}} $ should be delicately chosen on the event of  $ q \ne u $.


Regarding the LoS channel vector ${{\mathbf{z}}^t}$, each entry thereof is considered to have a non-stationary Gaussian prior distribution conditioned on a specified precision $ \rho _{m,u}^t $, i.e.,
\begin{equation}
p\left( {{{\mathbf{z}}^t}|{\bm{\rho }^t}} \right) = \prod\limits_{m,u} {p\left( {z_{{\text{R}},m,u}^t|\rho _{{\text{R}},m,u}^t} \right)} p\left( {z_{{\text{B}},m,u}^t|\rho _{{\text{B}},m,u}^t} \right),
\end{equation}
in which $ p\left( {z_{{\text{R}},m,u}^t|\rho _{{\text{R}},m,u}^t} \right) = \prod\nolimits_{m,u} {\mathcal{C}\mathcal{N}\left( {z_{{\text{R}},m,u}^t;0,\left( {1/\rho _{{\text{R}},m,u}^t} \right)} \right)}  $ and $ p\left( {z_{{\text{B}},m,u}^t|\rho _{{\text{B}},m,u}^t} \right)  = \prod\nolimits_{m,u}  {\mathcal{C}\mathcal{N}\left( {z_{{\text{B}},m,u}^t;0,\left( {1/\rho _{{\text{B}},m,u}^t} \right)} \right)}  $.

\subsubsection{Joint Distribution of the Layered Group Sparsity for LoS Channel}
As regards the LoS channel, the joint distribution of $ {{\bm{q}}^{1:t}} $, ${{\bm{\rho }}^{1:t}}$, and $ {{\mathbf{z}}^{1:t}} $ is given by
\begin{equation}
	p\left( {{{\bm{q}}^{1:t}},{{\mathbf{z}}^{1:t}},{{\bm{\rho }}^{1:t}}} \right) = \prod\limits_\tau  {\underbrace {p\left( {{{\bm{q}}^\tau }} \right)}_{{\text{Off-grid location}}}\underbrace {p\left( {{{\bm{\rho }}^\tau }|{{\bm{q}}^\tau }} \right)}_{{\text{Precision}}}} \underbrace {p\left( {{{\mathbf{z}}^\tau }|{{\bm{\rho }}^\tau }} \right)}_{{\text{LoS channel}}}.
\end{equation}

Note that the LoS channel vector $ {\bf{z}}^t $ can be segmented into $ UM $ blocks ($U$ blocks for each VUE $m$) with the $u$th block carrying the channel energies of both direct and cascaded channels, as both energies are all concentrated on the same location index $q_m^t$. More explicitly, the conditional prior in (\ref{layer-II}) imparts fewer precisions to the blocks fulfilling $ q=u $, forcing the entries thereof to depart from zero. For the opposite event of the blocks satisfying $q \neq u $, more precisions are assigned by the conditional prior in (\ref{layer-II}) to focus the block entries towards zero. Thus, the LoS channel vector contains separate blocks having distinct precisions for capturing different path gains, with the non-zero entry indexing the exact grid location of the $m$th VUE, termed as the \textit{group sparsity}.


\subsection{NLoS Channel}
We also scrutinize NLoS channel modeling at the BS/RIS using a layered heterogeneous paradigm. 
For the sake of clarity, we use the RIS-VUE link as an instance employing only a set of self-contained variables to construct hierarchical priors for the exact channel distributions{\footnote{Actually, there is still NLoS channel support over the BS-VUE link. The hierarchical priors of the BS-VUE NLoS channel continues to share with those presented in the RIS-VUE NLoS channel.}}. Thus in this instance, we just use ${\bf{v}}^t$ in place of $ {\bf{v}}_{\text{R}}^t $ or $ {\bf{v}}_{\text{B}}^t $ for notational brevity. Denote $ {\bm{c}^t} = {\left[ {\bm{c}_1^t;...;\bm{c}_M^t} \right]^T} \in {\mathbb{C}^{\tilde NM \times 1}},\forall t $ by the NLoS channel support, in which $ \bm{c}_m^t = {\left[ {\bm{c}_{m,1}^t,...,\bm{c}_{m,\tilde N}^t} \right]^T} \in {\mathbb{C}^{\tilde N \times 1}},\forall m,t $ with $ \bm{c}_{m,n}^t \in \left\{ {0,1} \right\} $ indicating whether there is an active NLoS path arriving at the BS from the $n$th angular grid at the RIS. Then, let $ {\bm{\gamma }^t} = {\left[ {\bm{\gamma }_1^t;...;\bm{\gamma }_M^t} \right]^T} \in {\mathbb{C}^{\tilde NM \times 1}},\forall t $ denote the precise vector of the NLoS channel support with $ \bm{\gamma }_m^t = {\left[ {\gamma _{m,1}^t,...,\gamma _{m,\tilde N}^t} \right]^T} \in {\mathbb{C}^{\tilde N \times 1}},\forall m,t $, for which $ \gamma _{m,n}^t $ represents the precision of $ v_{m,n}^t $, i.e., $ 1/\gamma _{m,n}^t $ is the variance of $ v_{m,n}^t $.

\subsubsection{Layer I: Markov Model for NLoS Channel Support}
In a time-varying environment, the AoAs at the BS/RIS may change slowly on the timescale of slots, and thus we employ the following model to capture the temporal correlation of the non-zero entries in the NLoS channel support~$\bm{c}^t$ 
\begin{equation}
c_{m,n}^t = {\tilde \varrho _{m,n}}c_{m,n}^{t - 1} + \sqrt {1 - \tilde \varrho _{m,n}^2} \mathcal{C}\mathcal{N}\left( {\alpha _{m,n}^t;0,\tilde \sigma _{m,n}^2} \right),
\end{equation}
where $ {\tilde {\varrho} _{m,k,n,q}} \in \left( {0,1} \right) $ characterizes the temporal correlation between two consecutive slots, termed as the temporal correlation coefficient, and $ \mathcal{C}\mathcal{N}\left( {\alpha _{m,n}^t;0,\tilde \sigma _{m,n}^2} \right) $ is the added Gaussian perturbation with zero mean and variance $ {\tilde \sigma _{m,n}^2} $. Hence, the NLoS channel support $\bm{c}^t$ takes the form of a conditional probability as follows
\begin{align}
p\left( {{\bm{c}^{1:t}}} \right) &= p\left( {{\bm{c}^1}} \right)\prod\limits_{\tau  = 2}^t {p\left( {{\bm{c}^\tau }|{\bm{c}^{\tau  - 1}}} \right)} \nonumber\\ &= \prod\limits_{m,n} {p\left( {c_{m,n}^1} \right)\prod\limits_{\tau  = 2}^t {p\left( {c_{m,n}^\tau |c_{m,n}^{\tau  - 1}} \right)} } .
\end{align}
Concerning the smooth dynamic behaviors inherent in the time-varying channels, we are allowed to identify the following independent first-order auto-regressive processes for the temporal correlation of the non-zero entries in $ \bm{c}^t $ \cite{AngDom-14} 
\begin{align}
	p\left( {c_{m,n}^\tau |c_{m,n}^{\tau  - 1}} \right) &= \prod\limits_{m,n} {p\left( {c_{m,n}^\tau |c_{m,n}^{\tau  - 1};{{\tilde {\varrho} }_{m,n}},{{\tilde {\sigma} }_{m,n}}} \right)} \nonumber\\
	 & = \prod\limits_{m,n} {\mathcal{C}\mathcal{N}\left( {c_{m,n}^\tau ;{{\tilde {\varrho} }_{m,n}}\alpha _{m,n}^{t - 1},\left( {1 - \tilde {\varrho }_{m,n}^2} \right)\tilde {\sigma} _{m,n}^2} \right)} ,
\end{align}
where $ p\left( {c_{m,n}^1} \right) \triangleq p\left( {c_{m,n}^1|c_{m,n}^0} \right) = \mathcal{C}\mathcal{N}\left( {c_{m,n}^1;0,\tilde \sigma _{m,n}^2} \right) $ specifies a steady-state distribution. 
To this end, the Markov model for the NLOS channel supports can be applicable for encapsulating various channel realizations in practice by subtly tuning parameters thereof. Hence, our analysis takes into account the steadily changing propagation environment between VUEs and BS/RIS, in which case the temporal coefficient ${\tilde \varrho _{m,n}}$ remains (almost) static during the tracking process of interest.

\subsubsection{Layers II \& III: A Pair of Conjugate Distributions}
The conditional prior of the precision vector $ {\bm{\gamma }^t} $  associated with the NLoS channel vector $ {{\mathbf{v}}^t} $  is given by
\begin{align}
p\left( {{\bm{\gamma }^t}|{\bm{c}^t}} \right) &= \prod\limits_{m,n} {p\left( {\gamma _{m,n}^t|c_{m,n}^t} \right)} \nonumber\\ &= \resizebox{0.78\hsize}{!}{$ \prod\limits_{m,n} {\Gamma {{\left( {\gamma _{m,n}^t;\mathfrak{a}_m^N,\mathfrak{b}_m^N} \right)}^{\mathbb{I}\left( {c_{m,n}^t} \right)}}\Gamma {{\left( {\gamma _{m,n}^t;\bar {\mathfrak{a}}_m^N,\bar {\mathfrak{b}}_m^N} \right)}^{\mathbb{I}\left( {1 - c_{m,n}^t} \right)}}} $}.
\end{align}
The case of  $ c_{m,n}^t = 1 $ indicates that there is an active NLoS path between the RIS and the $m$th VUE, in which the shape and rate parameters ${\mathfrak{a}}_m^N$ and $ {\mathfrak{b}}_m^N $ are required to satisfy $ \mathfrak{a}_m^N/{\mathfrak{b}}_m^N = \mathbb{E}\left[ {\gamma _{m,n}^t} \right],\forall m,n $  due to the fact that the variance $1/\gamma _{m,n}^t$ of $z_{m,u}^t$ equals to $\mathbb{E}\left[ {\gamma _{m,n}^t} \right]$ for the active NLoS path. With respect to the opposite case of $c_{m,n}^t = 0$, it holds that $\bar {\mathfrak{a}}_m^N/\bar {\mathfrak{b}}_m^N = \mathbb{E}\left[ {\gamma _{m,n}^t} \right] \gg 1,\forall m,n$ owing to the fact that the variance $ 1/\gamma _{m,n}^t $  approaches zero for the associated inactive NLoS path. 

As regards NLoS channels, a non-stationary Gaussian prior distribution with varying precision $\gamma _{m,n}^t$ is assigned to each entry of the NLoS channel ${{\mathbf{v}}^t}$ as follows
\begin{align}
p\left( {{{\mathbf{v}}^t}|{\bm{\gamma }^t}} \right) &= \prod\limits_{m,n} {p\left( {v_{m,n}^t|\gamma _{m,n}^t} \right)} \nonumber\\ &= \prod\limits_{m,n} {\mathcal{C}\mathcal{N}\left( {v_{m,n}^t;0,\left( {1/\gamma _{m,n}^t} \right)} \right)} .
\end{align}

\subsubsection{Joint Distribution of the Layered Group Sparsity for NLoS Channel}
Accordingly, the joint distribution of $\bm{c}^t$, $\bm{\gamma}$, and $\mathbf{v}$ is given by
\begin{equation}
p\left( {{\bm{c}^{1:t}},{\bm{\gamma }^{1:t}},{{\mathbf{v}}^{1:t}}} \right) = \prod\limits_\tau  {\underbrace {p\left( {{\bm{c}^\tau }} \right)}_{{\text{NLoS channel support}}}\underbrace {p\left( {{\bm{\gamma }^\tau }|{\bm{c}^\tau }} \right)}_{{\text{Precision}}}} \underbrace {p\left( { {\mathbf{v}}^\tau }| {\bm{\gamma}^\tau } \right)}_{{\text{NLoS channel}}}.
\end{equation}
Additionally, recall that the noise in (\ref{ori-signal}) follows a complex Gaussian distribution, and we denote $ \kappa ^t = \sigma ^{ - 2} $ by the noise precision. The noise precision $\kappa^t$ is typically unavailable and can be modeled using a Gamma distribution as a hyperprior $ p\left( {{\kappa }^t} \right) =  {\Gamma \left( {\kappa ^t;\mathfrak{a}^\kappa ,\mathfrak{b}^\kappa } \right)}  $, where the hyper-parameters $ \mathfrak{a}^\kappa  $ and $ \mathfrak{b}^\kappa  $ can be tuned towards near zero, i.e., $\mathfrak{a}^\kappa ,\mathfrak{b}^\kappa  \to 0$, for an approximation of a more general hyperprior \cite{AngDom-11}.

Significant angles typically appear in bursts (clusters), with each burst coinciding to a scattering cluster in the propagation environment, as contrasted to the group sparsity associated with the LoS channel vector ${\bf{z}}^t$ previously demonstrated. This is referred to as the \textit{burst sparsity}.


\subsection{Problem Formulation}

Our goal is to estimate the off-grid location offsets and AoA offset vectors associated with NLoS path, i.e., $ {{\mathbf{\hat \Omega }}^t} = \left\{ {\Delta {{{\mathbf{\hat r}}}^t},{{\left( {\Delta {{\bm{\hat {\omega} }}^{{\text{AoA}}}}} \right)}^t},{{\left( {\Delta {{\bm{\hat {\varphi} }}^{{\text{AoA}}}}} \right)}^t},{{\left( {\Delta { {\bm{\hat \vartheta} }^{{\text{AoA}}}}} \right)}^t}} \right\} $, given the observations up to the $ t $th slot ${{\mathbf{y}}^{1:t}} = \left[ {{\mathbf{y}}_1^{1:t};...;{\mathbf{y}}_M^{1:t}} \right] \in {\mathbb{C}^{KM \times 1}} $ in (\ref{ori-signal}) and $ {{\mathbf{\hat \Omega }}^{t - 1}} $ up to the $ (t-1) $th slot. To be specific, the offset parameters $ {{\mathbf{\Omega }}^t} $ can be obtained by a maximization of the likelihood function as follows 
\begin{align}\label{pro-ML}
	{{{\mathbf{\hat \Omega }}}^t} &= \mathop {\arg \max }\limits_{{{\mathbf{\Omega }}^t}} \ln p\left( {{{\mathbf{y}}^{1:t}};{{{\mathbf{\hat \Omega }}}^{1:t - 1}},{{\mathbf{\Omega }}^t}} \right) \nonumber\\
	 &= \mathop {\arg \max }\limits_{{{\mathbf{\Omega }}^t}} \int_{{\bm{\alpha }^t}} {\ln p\left( {{{\mathbf{y}}^{1:t}},{\bm{\alpha }^t};{{{\mathbf{\hat \Omega }}}^{1:t - 1}},{{\mathbf{\Omega }}^t}} \right)d{\bm{\alpha }^t}}, 
\end{align}
in which we denote ${\bm{\alpha }^t} = \left\{ {{{\mathbf{z}}^t},{{\mathbf{v}}^t},{\bm{\rho }^t},{\gamma ^t},{\bm{q}^t},{\bm{c}^t}} \right\}$ by the collection of corresponding variables for brevity. We attempt to determine the off-grid location index $ q_m^t $ by  calculating marginal posteriors $ p\left( {q_m^t|{{\mathbf{y}}^{1:t}};{{{\mathbf{\hat {\Omega} }}}^{1:t - 1}},{{\mathbf{\Omega }}^t}} \right) $ for the given estimates of offset parameters $ {{{\mathbf{\Omega }}^t}}$, after which the estimate $ \hat q_m^t  $ of $ {q_m^t} $ might be provided by the MAP probability estimate
\begin{equation}\label{MAP}
	\hat q_m^t = \mathop {\arg \max }  \limits_{q_m^t \in \left\{ {1,...,U} \right\}}  \ p\left( {q_m^t|{{\mathbf{y}}^{1:t}};{{{\mathbf{\hat \Omega }}}^{1:t - 1}},{{\mathbf{\Omega }}^t}} \right),
\end{equation}
with the conditional marginal posterior $p\left( {q_m^t|{{\mathbf{y}}^{1:t}};{{{\mathbf{\hat \Omega }}}^{1:t - 1}},{{\mathbf{\Omega }}^t}} \right)$ given by
\begin{align}\label{conditional-posterior}
	&p\left( {q_m^t|{{\mathbf{y}}^{1:t}};{{{\mathbf{\hat \Omega }}}^{1:t - 1}},{{\mathbf{\Omega }}^t}} \right) \nonumber\\ &\propto \sum\limits_{{\bm{c}^t}} {\int_{ - q_m^t,{{\mathbf{z}}^t},{{\mathbf{v}}^t},{\bm{\rho }^t},{\gamma ^t}} {p\left( {{{\mathbf{y}}^{1:t}},{\bm{\alpha }^t};{{{\mathbf{\hat \Omega }}}^{1:t - 1}},{{\mathbf{\Omega }}^t}} \right)} } \nonumber\\
	 &= \sum\limits_{{\bm{c}^t}} {\int_{ - q_m^t,{{\mathbf{z}}^t},{{\mathbf{v}}^t},{\bm{\rho }^t},{\gamma ^t}} {p\left( {{\bm{\alpha }^t};{{{\mathbf{\hat \Omega }}}^{1:t - 1}},{{\mathbf{\Omega }}^t}} \right)p\left( {{{\mathbf{y}}^{1:t}}|{\bm{\alpha }^t}} \right)} } ,
\end{align}
where $  - q_m^t $ denotes the vector collection over $ {\bm{q}^t} $  except for the entry $ q_m^t $, and $\propto$ refers to the equality after scaling. Consequently, the location estimate of the $m$th VUE in the $t$th slot can be determined by $ {\mathbf{\hat p}}_m^t = {\mathbf{r}}_{m,\hat q}^t + \Delta {\mathbf{\hat r}}_{m,\hat q}^t $.


Actually, it is quite challenging to attain the precise estimate of $ {\mathbf{\hat p}}_m^t, \forall m $, due to the prohibitive integration in (\ref{conditional-posterior}). More specifically, problem (\ref{conditional-posterior}) exposes the time-sequence attribute due to the spatial-temporal correlated off-grid location $\bm{q}^t$, which implies that the determination of $\bm{q}^t$ in each slot relies upon the observations ${\bf{y}}^{1:t}$ up to the current slot $t$. As $ t $ increases, the location tracking may become formidable owing to the excessively high computing complexity and memory costs. Furthermore, the complex priors of the RIS/BS-associated latent variables (such as $ \bm{c}^t $, $\bm{q}^t$, $\bm{\rho}^t$, $\bm{\gamma}^t$) and their intricate couplings (correlations) impart intractability to the exact conditional marginal posterior in (\ref{conditional-posterior}). This also induces the non-convexity in (\ref{pro-ML}) and renders some general optimization techniques fruitless for achieving a stationary solution. Therefore, the section that follows goes on to present beneficial techniques for clearing up the above obstacles, as well as an effective algorithm for the VUE location tracking with high accuracy and great robustness.

\begin{figure*}[ht]
	\centering
	\includegraphics[width=0.8\textwidth]{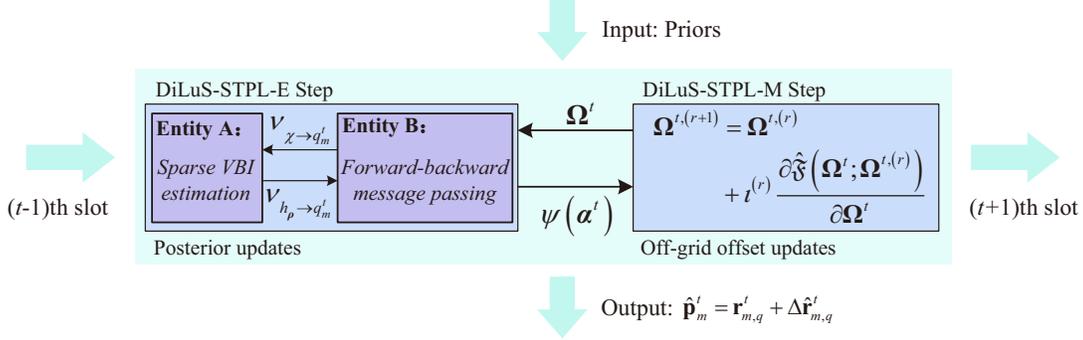}
	\caption{The proposed DiLuS-STPL algorithm.} \label{alg_block_diag}
\end{figure*}

\section{DiLuS-STPL Algorithm}\label{sec-DiLuS-STPL-Algorithm}

The fundamental goal of our examined location tracking is the robust and accurate recovery of the sparse LoS/NLoS channel vectors, e.g., $\mathbf{z}^t$ and $\mathbf{v}^t$, and the offset parameters $\mathbf{\Omega}^t$ based on the received signals $\mathbf{y}^{1:t}$ up to the $ t $th slot. Despite this, the computing complexity and memory costs involved should be taken into account if running an algorithm that performs tracking in each slot relying on the observations up to its current slot. To tackle this critical issue, this section aims to propose DiLuS-STPL algorithm in support of the intra-slot VBI estimates of $\mathbf{z}^t$, $\mathbf{v}^t$, and $\mathbf{\Omega}^t$, while incorporating the inter-slot message passing at a reduced complexity and overhead.

\subsection{Temporal Decomposition and Problem Approximation}

The joint probability distribution in~(\ref{pro-ML}) can be decomposed into two lightweight parts. In particular, one part covers the joint distribution in the $t$th slot which can be harnessed to determine $\mathbf{z}^t$, $\mathbf{v}^t$, and $\mathbf{\Omega}^t$. The other part, actually being the message passed from the previous slot, incorporates the information delivered by previous observations $\mathbf{y}^{1:t-1}$.
Firstly, we restructure the joint distribution in the $t$th slot $p\left( {{{\mathbf{y}}^t},{\bm{\alpha }^t};{{\mathbf{\Omega }}^t}} \right)$ as
\begin{align}
 p\left( {{{\mathbf{y}}^t},{\bm{\alpha }^t};{{\mathbf{\Omega }}^t}} \right) =& p\left( {{{\mathbf{y}}^t},{{\mathbf{z}}^t},{{\mathbf{v}}^t},{\bm{\rho }^t},{\bm{\gamma} ^t},{\bm{q}^t},{\bm{c}^t}} \right) \nonumber\\
\mathop  = \limits^{\left( a \right)} & p\left( {{{\mathbf{y}}^t}|{{\mathbf{z}}^t},{{\mathbf{v}}^t};{{\mathbf{\Omega }}^t}} \right)p\left( {{{\mathbf{z}}^t}|{\bm{\rho }^t}} \right) p\left( {{\bm{\rho }^t}|{\bm{q}^t}} \right) \nonumber\\ 
& \times p\left( {{{\mathbf{v}}^t}|{\bm{\gamma} ^t}} \right)p\left( {{\bm{\gamma} ^t}|{\bm{c}^t}} \right)p\left( {{\bm{q}^t}} \right)p\left( {{\bm{c}^t}} \right),
\end{align}
in which (a) holds due to the proposed DiLuS framework. By taking into account the priors from the previous slot, the joint distribution in (\ref{pro-ML}) can be recast leveraging the sequential Bayesian inference technique \cite{Bayesian-3,book-BMTT}, i.e.,
\begin{align}\label{ML-tmp}
&p\left( {{{\mathbf{y}}^{1:t}},{\bm{\alpha }^t};{{{\mathbf{\hat {\Omega} }}}^{1:t - 1}},{{\mathbf{\Omega }}^t}} \right) \nonumber\\ & \mathop  \propto \limits^{\left( b \right)} p\left( {{{\mathbf{y}}^t},{\bm{\alpha }^t};{{\mathbf{\Omega }}^t}} \right)p\left( {{\bm{\alpha }^t}|{{\mathbf{y}}^{1:t - 1}};{{{\mathbf{\hat {\Omega} }}}^{1:t - 1}}} \right),
\end{align}
where (b) holds due to \cite[Eq. (3.2)]{Bayesian-3}. The conditional posterior $ p\left( {{\bm{\alpha }^t}|{{\mathbf{y}}^{1:t - 1}};{{{\mathbf{\hat \Omega }}}^{1:t - 1}}} \right) $ is given by
\begin{align}
	&p\left( {{\bm{\alpha }^t}|{{\mathbf{y}}^{1:t - 1}};{{{\mathbf{\hat \Omega }}}^{1:t - 1}}} \right) \nonumber\\& \mathop  = \limits^{\left( c \right)} \int {\underbrace {p\left( {{\bm{\alpha }^t}|{\bm{\alpha }^{t - 1}}} \right)}_{{\text{Transition}}}\underbrace {p\left( {{\bm{\alpha }^{t - 1}}|{{\mathbf{y}}^{1:t - 1}};{{{\mathbf{\hat \Omega }}}^{1:t - 1}}} \right)}_{{\text{Prior}}}d{\bm{\alpha }^{t - 1}}} ,
\end{align}
where (c) holds due to \cite[Eq. (3.12)]{book-BMTT} with $ {p\left( {{\bm{\alpha }^t}|{\bm{\alpha }^{t - 1}}} \right)} $ given by 
\begin{equation}
p\left( {{\bm{\alpha }^t}|{\bm{\alpha }^{t - 1}}} \right) = p\left( {{{\mathbf{z}}^t},{{\mathbf{v}}^t},{\bm{\rho }^t},{\bm{\gamma} ^t}|{\bm{q}^t},{\bm{c}^t}} \right)p\left( {{\bm{q}^t}|{\bm{q}^{t - 1}}} \right)p\left( {{\bm{c}^t}|{\bm{c}^{t - 1}}} \right).
\end{equation}
Regarding the conditional posterior $ p\left( {{\bm{\alpha }^t}|{{\mathbf{y}}^{1:t - 1}};{{{\mathbf{\hat \Omega }}}^{1:t - 1}}} \right) $, we have 
\begin{align}
&p\left( {{\bm{\alpha }^t}|{{\mathbf{y}}^{1:t - 1}};{{{\mathbf{\hat \Omega }}}^{1:t - 1}}} \right) \nonumber\\ &\mathop  = \limits^{\left( d \right)}  p\left( {{{\mathbf{z}}^t},{{\mathbf{v}}^t},{\bm{\rho }^t},{\gamma ^t}|{\bm{q}^t},{\bm{c}^t}} \right)\sum\limits_{{\bm{q}^{t - 1}},{\bm{c}^{t - 1}}} {p\left( {{\bm{q}^t}|{\bm{q}^{t - 1}}} \right)p\left( {{\bm{c}^t}|{\bm{c}^{t - 1}}} \right)} \nonumber\\ & \quad \times  p\left( {{\bm{q}^{t - 1}},{\bm{c}^{t - 1}}|{{\mathbf{y}}^{1:t - 1}};{{{\mathbf{\hat \Omega }}}^{1:t - 1}}} \right),
\end{align}
for which (d) holds due to \cite[Eq. (3.12)]{book-BMTT}. Hence, $ p\left( {{{\mathbf{y}}^{1:t}},{\bm{\alpha }^t};{{{\mathbf{\hat {\Omega} }}}^{1:t- 1}},{{\mathbf{\Omega }}^t}} \right) $ in (\ref{ML-tmp}) can be further structured as
\begin{align}\label{prop}
&p\left( {{{\mathbf{y}}^{1:t}},{\bm{\alpha }^t};{{{\mathbf{\hat \Omega }}}^{1:t - 1}},{{\mathbf{\Omega }}^t}} \right) \nonumber\\ 
&\propto  \ p\left( {{{\mathbf{y}}^t},{\bm{\alpha }^t};{{\mathbf{\Omega }}^t}} \right)p\left( {{\bm{\alpha }^t}|{{\mathbf{y}}^{1:t - 1}};{{{\mathbf{\hat \Omega }}}^{1:t - 1}}} \right) \nonumber\\
 &= \ p\left( {{{\mathbf{y}}^t}|{{\mathbf{z}}^t},{{\mathbf{v}}^t};{{\mathbf{\Omega }}^t}} \right)p\left( {{{\mathbf{z}}^t}|{\bm{\rho }^t}} \right)p\left( {{\bm{\rho }^t}|{\bm{q}^t}} \right)p\left( {{{\mathbf{v}}^t}|{\bm{\gamma} ^t}} \right)p\left( {{\bm{\gamma} ^t}|{\bm{c}^t}} \right) \nonumber\\
  & \quad \times \sum\limits_{{\bm{q}^{t - 1}},{\bm{c}^{t - 1}}} p\left( {{\bm{q}^t}|{\bm{q}^{t - 1}}} \right)p\left( {{\bm{c}^t}|{\bm{c}^{t - 1}}} \right) \nonumber\\ 
  	& \quad \times p\left( {{\bm{q}^{t - 1}},{\bm{c}^{t - 1}}|{{\mathbf{y}}^{1:t - 1}};{{{\mathbf{\hat \Omega }}}^{1:t - 1}}} \right) .
\end{align}
Owing to the intractability of the exact posterior $ p\left( {{\bm{q}^{t - 1}},{\bm{c}^{t - 1}}|{{\mathbf{y}}^{1:t - 1}};{{{\mathbf{\hat \Omega }}}^{1:t - 1}}} \right) $, one may need a safe approximation in its place, i.e., $ p\left( {{\bm{q}^{t - 1}},{\bm{c}^{t - 1}}|{{\mathbf{y}}^{1:t - 1}};{{{\mathbf{\hat \Omega }}}^{1:t - 1}}} \right) \approx \psi \left( {{\bm{q}^{t - 1}}} \right)\psi \left( {{\bm{c}^{t - 1}}} \right) $ with $ \psi \left( {{\bm{q}^{t - 1}}} \right) \approx p\left( {{\bm{q}^{t - 1}}|{{\mathbf{y}}^{1:t - 1}};{{{\mathbf{\hat \Omega }}}^{1:t - 1}}} \right) $ and $ \psi \left( {{\bm{c}^{t - 1}}} \right) \approx p\left( {{\bm{c}^{t - 1}}|{{\mathbf{y}}^{1:t - 1}};{{{\mathbf{\hat \Omega }}}^{1:t - 1}}} \right) $. In this case, the joint prior distribution in (\ref{prop}) can be further recast as (\ref{approx-joint}) shown at the top of the next page,
\begin{figure*}[!htb]
\begin{align}\label{approx-joint}
	&p\left( {{{\mathbf{y}}^{1:t}},{\bm{\alpha }^t};{{{\mathbf{\hat \Omega }}}^{1:t - 1}},{{\mathbf{\Omega }}^t}} \right) \approx \hat p\left( {{{\mathbf{y}}^{1:t}},{\bm{\alpha }^t};{{{\mathbf{\hat \Omega }}}^{1:t - 1}},{{\mathbf{\Omega }}^t}} \right) \nonumber\\
	 &= p\left( {{{\mathbf{y}}^t}|{{\mathbf{z}}^t},{{\mathbf{v}}^t};{{\mathbf{\Omega }}^t}} \right)p\left( {{{\mathbf{z}}^t}|{\bm{\rho }^t}} \right)p\left( {{\bm{\rho }^t}|{\bm{q}^t}} \right)p\left( {{{\mathbf{v}}^t}|{\bm{\gamma} ^t}} \right)p\left( {{\bm{\gamma} ^t}|{\bm{c}^t}} \right) 
	  \sum\limits_{{\bm{q}^{t - 1}},{\bm{c}^{t - 1}}} {p\left( {{\bm{q}^t}|{\bm{q}^{t - 1}}} \right)p\left( {{\bm{c}^t}|{\bm{c}^{t - 1}}} \right)\psi \left( {{\bm{q}^{t - 1}}} \right)\psi \left( {{\bm{c}^{t - 1}}} \right)} \nonumber\\
	   &= p\left( {{{\mathbf{y}}^t}|{{\mathbf{z}}^t},{{\mathbf{v}}^t};{{\mathbf{\Omega }}^t}} \right)p\left( {{{\mathbf{z}}^t}|{\bm{\rho }^t}} \right)p\left( {{\bm{\rho }^t}|{\bm{q}^t}} \right)p\left( {{{\mathbf{v}}^t}|{\bm{\gamma} ^t}} \right)p\left( {{\bm{\gamma} ^t}|{\bm{c}^t}} \right)\hat p\left( {{\bm{q}^t}} \right)\hat p\left( {{\bm{c}^t}} \right),
\end{align}
\hrule
\end{figure*}
where the approximate priors of $\bm{q}^t$ and  $\bm{c}^t$ are particularly given by $\hat {p}\left( {{\bm{q}^t}} \right) \triangleq \sum\nolimits_{{\bm{q}^{t - 1}}} {p\left( {{\bm{q}^t}|{\bm{q}^{t - 1}}} \right)\psi \left( {{\bm{q}^{t - 1}}} \right)} $ and $\hat {p}\left( {{\bm{c}^t}} \right) \triangleq \sum\nolimits_{{\bm{c}^{t - 1}}} {p\left( {{\bm{c}^t}|{\bm{c}^{t - 1}}} \right)\psi \left( {{\bm{c}^{t - 1}}} \right)} $. Note that the only discrepancy between the prior in (\ref{approx-joint}) and the original prior lies in that the dynamic hidden Markov model  $ p\left( {{\bm{q}^t}} \right) $ associated with the off-grid location is replaced by an approximate prior  $ \hat p\left( {{\bm{q}^t}} \right) $ with independent entries, while the Markov model $p\left( {{\bm{c}^t}} \right)$ is approximated by its safe copy  $ \hat p\left( {{\bm{c}^t}} \right) $. In what follows, we expound the realization of the approximate posteriors $ \psi \left( {{\bm{q}^{t - 1}}} \right) $ and $\psi \left( {{\bm{c}^{t - 1}}} \right)$.

\subsection{Outline of DiLuS-STPL}
As a key feature of DiLuS-STPL, the approximate posteriori $ \psi \left( {{\bm{\alpha }^t}|{{\mathbf{y}}^{1:t}}; {{\mathbf{\Omega }}^t}} \right) $ can be attained in the DiLuS-STPL-E Step using the observation at the $ t $th slot and the message passed from the previous slot, while updating ${{\mathbf{\Omega }}^t}$ by leveraging a gradient-based technique, e.g., majorization-minimization (MM), in the DiLuS-STPL-M Step. Due to myriads of loops in such a dense factor graph, the classical sum-product message passing (SPMP) \cite{TIT-2001} fails to traverse the factor and variable nodes over the entire factor graph. As a potential resolution, the approximate message passing (AMP)-based algorithms, e.g., Turbo-AMP \cite{AngDom-11-11} and Turbo-CS \cite{AngDom-11-12}, tend to be entrapped into their local optimum and exhibit usually poor performance. In an effort to facilitate the practical implementation, two entities are specified in the DiLuS-STPL-E Step, i.e., Entity A and Entity B as illustrated in Fig.~\ref{alg_block_diag}, in support of sparse VBI estimates and forward-backward message passing, such that the proposed algorithm may achieve approximate message passing over the entire factor graph with a faithful performance. Concretely, the message passing over the entire factor graph can be interpreted as an iteration of this pair of partitioned entities, one of which undertakes inference on one subgraph that encapsulates the group sparsity in the observation, and the other of which plays inference on the subgraph that captures the spatial correlation among vehicle positions. Furthermore, the interactions between the twin entities may effectively eliminate the self-reinforcement during the update procedure.


\subsubsection{Entity A}
This entity intends to obtain the approximate conditional marginal posteriors of latent variables using sparse VBI estimation, i.e., $ \psi \left( {{\bm{\alpha }^t}} \right) = \psi \left( {{\bm{\alpha }^t}|{{\mathbf{y}}^{1:t}};{{\mathbf{\Omega }}^t}} \right) $. Let $ {\nu _{\chi  \to q_m^t}}\left( {q_m^t} \right),m = 1,...,M $ denote the input of Entity A, which can be interpreted as the message passed from Entity~B and  incorporates the spatial correlation among different vehicles in the vehicle platoon. The sparse VBI estimates executed in Entity~A will be detailed later in Sec.~IV-D.

\subsubsection{Entity B}
\begin{figure*}[t]
	\centering
	\includegraphics[width=0.95\textwidth]{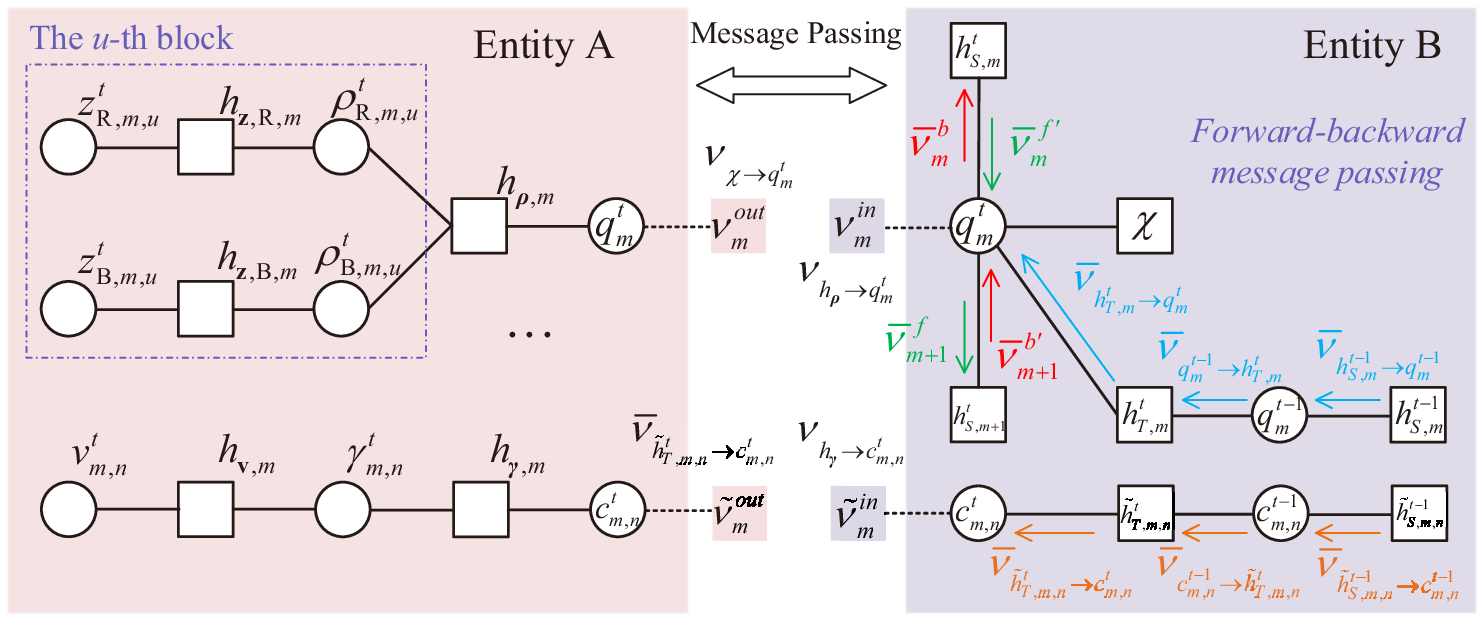}
	\caption{The partial factor subgraph and the interaction between the two entities. The factors are defined as follows: $ {h_{{\mathbf{z}},{\text{R}},m}} \triangleq p\left( {z_{{\text{R}},m,u}^t|\rho _{{\text{R}},m,u}^t} \right) $, $ {h_{{\mathbf{z}},{\text{B}},m}} \triangleq p\left( {z_{{\text{B}},m,u}^t|\rho _{{\text{B}},m,u}^t} \right) $, $ {h_{\bm{\rho },m}} \triangleq p\left( {\rho _{{\text{R}},m,u}^t|q_m^t} \right)p\left( {\rho _{{\text{B}},m,u}^t|q_m^t} \right) $, $ {h_{{\mathbf{v}},m}} \triangleq p\left( {v_{m,n}^t|\gamma _{m,n}^t} \right) $, and $ {h_{\bm{\gamma },m}} \triangleq p\left( {\gamma _{m,n}^t|c_{m,n}^t} \right) $.} \label{subFG}
\end{figure*}

The message $ {\nu _{{h_{\bm{\rho }}} \to q_m^t}}\left(  \cdot  \right)$  provided by Entity A constructs the input of Entity~B. For ease of exposition, the input and the output of~Entity B are defined as $ \nu _m^{in} \triangleq {\nu _{{h_{\bm{\rho }}} \to q_m^t}}\left(  \cdot  \right) $ and $ \nu _m^{out} \triangleq {\nu _{\chi  \to q_m^t}}\left(  \cdot  \right),m = 1,...,M$, respectively. In each algorithm iteration, the message $ \nu_m^{in}\left( {q_m^t} \right) = {\nu _{{h_{\bm{\rho }}} \to q_m^t}}\left( {q_m^t} \right) $ can be obtained in accordance with the belief propagation rule \cite{TIT-2001}
\begin{equation}
	\nu _m^{in}\left( {q_m^t} \right) \propto \frac{{\psi (q_m^t)}}{{\nu _m^{out} \left( {q_m^t} \right)}},\forall q_m^t.
\end{equation}
Next, let us examine the off-grid location $q_m^t$'s corresponding technique as an example of detailing the forward-backward message passing. By leveraging the belief propagation rule, the forward messages $ \bar {\nu} _m^{f^ \prime } $ and $ \bar {\nu} _m^f $ associated with $ q_m^t $ are respectively given by
\begin{align}
&\bar {\nu} _m^{f^ \prime }\left( {q_m^t} \right) \propto \left\{ \begin{gathered}
	p\left( {q_1^t} \right),\qquad \qquad \qquad \qquad \quad \ \, m = 1,  \hfill \\
	\int_{ - q_m^t} {p\left( {q_m^t|q_{m - 1}^t} \right)\bar {\nu} _m^f\left( {q_m^{t - 1}} \right),2 \le m \le M},  \hfill \\ 
\end{gathered}  \right. \\
&\bar \nu _{m + 1}^f\left( {q_m^t} \right) \propto \bar \nu _m^{f^ \prime }\left( {q_m^t} \right)\bar \nu _m^{in}\left( {q_m^t} \right){\bar \nu _{h_{T,m}^t \to q_m^t}}\left( {q_m^t} \right).
\end{align}
The message $ {\bar \nu _{h_{T,m}^t \to q_m^t}}\left( {q_m^t} \right) $ characterizes the temporal correlation of the $n$th VUE, implying that the location in the previous slot provides a priori for the current slot, which, in particular, is given~by
\begin{equation}\label{35}
{\bar \nu _{h_{T,m}^t \to q_m^t}}\left( {q_m^t} \right) \propto \int_{ - q_m^t} {p\left( {q_m^t|q_m^{t - 1}} \right){{\bar \nu }_{q_m^{t - 1} \to h_{T,m}^t}}\left( {q_m^{t - 1}} \right)} .
\end{equation}
In (\ref{35}), as demonstrated in Fig.~\ref{subFG}, message ${\bar \nu _{q_m^{t - 1} \to h_{T,m}^t}}\left( {q_m^{t - 1}} \right)$ corresponds to the approximate posterior $ \psi \left( {q_m^{t - 1}} \right) $ associated with the off-grid location $q_m^t$ of the $m$th VUE in the $t$th slot, and thus we have
\begin{equation}\label{approx-previous-slot}
	{\bar \nu _{q_m^{t - 1} \to h_{T,m}^t}}\left( {q_m^{t - 1}} \right) \propto {\bar \nu _{h_{S,m}^{t - 1} \to q_m^{t - 1}}}\left( {q_m^{t - 1}} \right) \triangleq \psi \left( {q_m^{t - 1}} \right).
\end{equation}
The backward messages $\bar \nu _m^{b^ \prime }$ and $\bar \nu _m^b$ respectively given by
\begin{align}
	\resizebox{1.0\hsize}{!}{$
	\bar \nu _{m + 1}^{b^ \prime }\left( {q_m^t} \right) \propto \int_{ - q_m^t} {p\left( {q_{m + 1}^t|q_m^t} \right)\bar \nu _{m + 1}^b\left( {q_{m + 1}^t} \right)} ,1 \le m \le M - 1,
	$}
\end{align}
\begin{align}
	&\bar \nu _m^b\left( {q_m^t} \right) \nonumber\\& \propto \left\{ \begin{gathered}
		\bar \nu _{m + 1}^{b^ \prime }\left( {q_m^t} \right)\bar \nu _m^{in}\left( {q_m^t} \right){{\bar \nu }_{h_{T,m}^t \to q_m^t}}\left( {q_m^t} \right),1 \le m \le M - 1, \hfill \\
		\bar \nu _M^{in}\left( {q_m^t} \right){{\bar \nu }_{h_{T,m}^t \to q_M^t}}\left( {q_m^t} \right), \qquad \qquad \ \ m = M . \hfill \\ 
	\end{gathered}  \right.
\end{align}
We omit here, for the sake of conciseness, the intricacies of the similar procedure that pertains to the message passing associated with the NLoS channel support $c_{m,n}^t$.
	
Having established the method for carrying out message passing between the twin entities, let us proceed to a thorough investigation of the DiLuS-STPL-M Step and the DiLuS-STPL-E Step in what follows, respectively, based upon which the off-grid location and corresponding parameters can be determined.

\subsection{DiLuS-STPL-M Step: Inexact~Majorization-Minimization~(MM)}
The likelihood function $\ln p\left( {{{\mathbf{y}}^{1:t}};{{\mathbf{\Omega }}^t}} \right)$ is indeed intractable owing to the absence of the closed-form expressions induced by the prohibitive integrals of the joint distribution with respect to $  {\bm{\alpha }^t}  $. An alternative is to construct its surrogate function based on $ {{\mathbf{\Omega }}^{t,(r)}} $ available in the $r$th iteration in place of the objective in problem (\ref{pro-ML}), providing that the exact posterior $ p\left( {{\bm{\alpha }^t}|{{\mathbf{y}}^{1:t}};{{\mathbf{\Omega }}^{t,\left( r \right)}}} \right) $ has been determined.
Unfortunately, the exact posterior $ p\left( {{\bm{\alpha }^t}|{{\mathbf{y}}^{1:t}};{{\mathbf{\Omega }}^{t,\left( r \right)}}} \right) $  is also highly intractable in our investigated problem because of the intricate loops intrinsic in the factor graph. To circumvent this issue, the VBI and message passing techniques are employed to identify an alternative $ \psi \left( {{\bm{\alpha }^t}} \right) $ for a faithful approximation of the posterior $ p\left( {{\bm{\alpha }^t}|{{\mathbf{y}}^{1:t}};{{\mathbf{\Omega }}^{t,\left( r \right)}}} \right) $. The to-be-identified posterior $ \psi \left( {{\bm{\alpha }^t}} \right) $ has a factorized form, i.e., $ \psi \left( {{\bm{\alpha }^t}} \right) = \prod\nolimits_{l \in \mathcal{L}} {\psi \left( {\bm{\alpha }_l^t} \right)}  $, in which $ \bm{\alpha }_l^t $ denotes an individual in $ {\bm{\alpha }^t} = \left\{ {{{\mathbf{z}}^t},{{\mathbf{v}}^t},{\bm{\rho }^t},{\bm{\gamma} ^t},{\bm{q}^t},{\bm{c}^t}} \right\} $, and we have $ \mathcal{L} = \left\{ {l|\forall \bm{\alpha }_l^t \in {\bm{\alpha }^t}} \right\} $. Based upon the posterior approximation, an equivalent of the original MLE problem in (\ref{pro-ML}) according to the following proposition.

\textit{\textbf{Proposition 1:}} Denote  $  {\mathbf{\Omega }}^{1:t,(r)}  $ by the optimal estimates achieved for problem (\ref{pro-ML}) in the $r$th iteration, the optimality of which is equivalent to the following optimization problem at the $t$th~slot,
\begin{equation}\label{approx-1}
\mathop {\max }\limits_{\psi \left( {{\bm{\alpha }^t}} \right),{{\mathbf{\Omega }}^t}} \  \underbrace {\int {\psi \left( {{\bm{\alpha }^t}} \right)\ln \frac{{p\left( {{{\mathbf{y}}^{1:t}},{\bm{\alpha }^t};{{{\mathbf{\hat \Omega }}}^{1:t - 1,\left( r \right)}},{{\mathbf{\Omega }}^t}} \right)}}{{\psi \left( {{\bm{\alpha }^t}} \right)}}d{\bm{\alpha }^t}} }_{\mathcal{F}\left( {{{\mathbf{\Omega }}^t};{{\mathbf{\Omega }}^{t,\left( r \right)}}} \right)}.
\end{equation}

\textit{Proof:} As regards the optimization problem in (\ref{approx-1}), we have 
\begin{align}
	&\int {\psi \left( {{\bm{\alpha }^t}} \right)\ln \frac{{p\left( {{{\mathbf{y}}^{1:t}},{\bm{\alpha }^t};{{{\mathbf{\hat \Omega }}}^{1:t - 1,\left( r \right)}},{{\mathbf{\Omega }}^t}} \right)}}{{\psi \left( {{\bm{\alpha }^t}} \right)}}d{\bm{\alpha }^t}} \nonumber\\ &\mathop  \le \limits^{\left( a \right)} \ln \int {\psi \left( {{\bm{\alpha }^t}} \right)\frac{{p\left( {{{\mathbf{y}}^{1:t}},{\bm{\alpha }^t};{{{\mathbf{\hat \Omega }}}^{1:t - 1,\left( r \right)}},{{\mathbf{\Omega }}^t}} \right)}}{{\psi \left( {{\bm{\alpha }^t}} \right)}}d{\bm{\alpha }^t}} \nonumber\\
	 &= \ln \int {p\left( {{{\mathbf{y}}^{1:t}},{\bm{\alpha }^t};{{{\mathbf{\hat \Omega }}}^{1:t - 1,\left( r \right)}},{{\mathbf{\Omega }}^t}} \right)d{\bm{\alpha }^t}} ,
\end{align}
where (a) holds due to the Jensen’s inequality. Thus, problem~(\ref{approx-1}) can be maximized when  $ \psi \left( {{\bm{\alpha }^t}} \right) $ has the following form 
\begin{equation}
	{\psi ^*}\left( {{\bm{\alpha }^t}} \right) = p\left( {{\bm{\alpha }^t}|{{\mathbf{y}}^{1:t}};{{{\mathbf{\hat \Omega }}}^{1:t - 1,\left( r \right)}},{{\mathbf{\Omega }}^t}} \right),
\end{equation}
and the optimization problem is pruned to 
\begin{equation}
	\mathop {\max } \limits_{{{\mathbf{\Omega }}^t}} \ \ln p\left( {{{\mathbf{y}}^{1:t}};{{{\mathbf{\hat \Omega }}}^{1:t - 1,\left( r \right)}},{{\mathbf{\Omega }}^t}} \right).
\end{equation}
Therefore, the optimal estimates, i.e., $ {\mathbf{\Omega }}^{t,(r)} $, are the optimum to the original problem~(\ref{approx-1}), which thus completes the proof.~\hfill$ \blacksquare  $

With the approximation of the joint distribution in (\ref{approx-joint}), the objective function in (\ref{approx-1}) can be recast as follows
\begin{equation}\label{surrogate-function}
	\hat {\mathcal{F}}\left( {{{\mathbf{\Omega }}^t};{{\mathbf{\Omega }}^{t,\left( r \right)}}} \right) = \int {\psi \left( {{\bm{\alpha }^t}} \right)\ln \frac{{\hat p\left( {{{\mathbf{y}}^{1:t}},{\bm{\alpha }^t};{{\mathbf{\hat \Omega }}}^{1:t - 1,\left( r \right)},{{\mathbf{\Omega }}^t}} \right)}}{{\psi \left( {{\bm{\alpha }^t}} \right)}}d{\bm{\alpha }^t}} .
\end{equation}
Then in the $ (r+1) $th iteration in the DiLuS-STPL-M Step, we update ${{\mathbf{\Omega }}^{t,\left( {r + 1} \right)}}$ as
\begin{equation}\label{tmp-44}
	{{\mathbf{\Omega }}^{t,\left( {r + 1} \right)}} = \mathop {\arg \max }\limits_{{{\mathbf{\Omega }}^t}} \ \hat {\mathcal{F}}\left( {{{\mathbf{\Omega }}^t};{{\mathbf{\Omega }}^{t,\left( r \right)}}} \right).
\end{equation}
The non-concave nature of the objective, nevertheless, impedes the pursuit of the global optimum for (\ref{tmp-44}). We thus exploit the following gradient update  in an effort to obtain a stationary solution
\begin{equation}\label{update-Omega}
	{{\mathbf{\Omega }}^{t,\left( {r + 1} \right)}} = {{\mathbf{\Omega }}^{t,\left( r \right)}} + {\iota ^{\left( r \right)}}\frac{{\partial \hat {\mathcal{F}}\left( {{{\mathbf{\Omega }}^t};{{\mathbf{\Omega }}^{t,\left( r \right)}}} \right)}}{{\partial {{\mathbf{\Omega }}^t}}},
\end{equation}
where ${\iota ^{\left( r \right)}}$ is the step size determined by the Armijo rule \cite{AngDom-11}.

\subsection{DiLuS-STPL-E Step: Closed-form Updates}
\subsubsection{Outline of Sparse VBI}
The approximate conditional marginal posterior can be determined by a minimization of the Kullback-Leibler divergence (KLD) between $ \hat p\left( {{{\mathbf{y}}^{1:t}},{\bm{\alpha }^t};{{{\mathbf{\hat \Omega }}}^{1:t - 1}},{{\mathbf{\Omega }}^t}} \right) $ and  $ \psi \left( {{\bm{\alpha }^t}} \right) $, subject to the factorized structure on $ \psi \left( {{\bm{\alpha }^t}} \right) $, thus yielding the following problem
\begin{align}\label{P1}
	{\psi ^*}\left( {{\bm{\alpha }^t}} \right) =& \mathop {\arg \min }\limits_{\psi \left( {{\bm{\alpha }^t}} \right),{{\mathbf{\Omega }}^t}} \int {\psi \left( {{\bm{\alpha }^t}} \right)\ln \frac{{\psi \left( {{\bm{\alpha }^t}} \right)}}{{\hat p\left( {{\bm{\alpha }^t}|{{\mathbf{y}}^{1:t}},;{{\mathbf{\Omega }}^t}} \right)}}} d{\bm{\alpha }^t} \nonumber\\
	&s.t. \quad \psi \left( {{\bm{\alpha }^t}} \right) = \prod\limits_{l \in \mathcal{L}} {\psi \left( {\bm{\alpha }_l^t} \right)} .
\end{align}
The stationary solution $ {{\psi ^*}\left( {\bm{\alpha }_l^t} \right)} $ to problem (\ref{P1}) needs to satisfy
\begin{align}
	&{\psi ^*}\left( {\bm{\alpha }_l^t} \right) \nonumber\\ 
	& \resizebox{1.0\hsize}{!}{$
	= \mathop {\arg \min }\limits_{\psi \left( {\bm{\alpha }_l^t} \right),{{\mathbf{\Omega }}^t}} \int {\prod\limits_{l \ne j} {{\psi ^*}\left( {\bm{\alpha }_l^t} \right)\psi \left( {\bm{\alpha }_j^t} \right)} \ln \frac{{\prod\nolimits_{l \ne j} {{\psi ^*}\left( {\bm{\alpha }_l^t} \right)\psi \left( {\bm{\alpha }_j^t} \right)} }}{{\hat p\left( {{\bm{\alpha }^t}|{{\mathbf{y}}^{1:t}};{{\mathbf{\Omega }}^t}} \right)}}} d{\bm{\alpha }^t}, 
	$}
\end{align}
which can be attained by triggering alternating updates of $ \psi \left( {\bm{\alpha }_l^t} \right),\forall l \in \mathcal{L} $. Specifically, for given $ \psi \left( {\bm{\alpha }_j^t} \right),j \ne l $, a unique solution that facilitates the KLD minimization in problem (\ref{P1}) is given by
\begin{equation}
	{\psi ^*}\left( {\bm{\alpha }_l^t} \right) \propto \left\{ \begin{gathered}
		\exp \left( {{\mathbb{E}_{\prod\nolimits_{l \ne j} {\psi \left( {\bm{\alpha }_j^t} \right)} }}\left[ {p\left( {{\bm{\alpha }^t}|{{\mathbf{y}}^{1:t}};{{\mathbf{\Omega }}^t}} \right)} \right]} \right),t = 1 ,\hfill \\
		\exp \left( {{\mathbb{E}_{\prod\nolimits_{l \ne j} {\psi \left( {\bm{\alpha }_j^t} \right)} }}\left[ {\hat p\left( {{\bm{\alpha }^t}|{{\mathbf{y}}^{1:t}};{{\mathbf{\Omega }}^t}} \right)} \right]} \right),t > 1, \hfill \\ 
	\end{gathered}  \right.
\end{equation}
where $ {\mathbb{E}_{h\left( x \right)}}\left[ {f\left( x \right)} \right] = \int {f\left( x \right)h\left( x \right)dx}  $.

\subsubsection{Initialization of Sparse VBI}
The sparse VBI implemented in Entity A is highly dependent on an appropriate initialization, based upon which an alternating algorithm is triggered for the
initialization of the corresponding posteriors. In the first outer iteration, the approximate prior of the off-grid location $\bm{q}^t$ is initialized by 
\begin{equation}\label{init-q}
\hat p\left( {{\bm{q}^t}} \right) = \prod\limits_m {\sum\limits_u {\pi _{m,u}^{\text{L}}\delta \left( {q_m^t - u} \right)} } ,
\end{equation}
where $ \pi _{m,u}^{\text{L}} = {\nu _{\chi  \to q_m^t}}\left( {q_m^t} \right) $. Additionally, the approximate prior of the NLoS channel support $\bm{c}^t$ can be given by 
\begin{equation}\label{init-c}
	\hat p\left( {{\bm{c}^t}} \right) = \sum\limits_{{\bm{c}^{t - 1}}} {p\left( {{\bm{c}^t}|{\bm{c}^{t - 1}}} \right)\psi \left( {{\bm{c}^{t - 1}}} \right)} .
\end{equation}

\subsubsection{Update Posterior for LoS Channels  $ \mathbf{z}^t $}
The LoS channel $\psi \left( {{{\mathbf{z}}^t}} \right)$’s can be updated by a complex Gaussian distribution with determined mean and variance 
\begin{equation}\label{update-z}
	\psi \left( {{{\mathbf{z}}^t}} \right) = \mathcal{C}\mathcal{N}\left( {{{\mathbf{z}_{\text{R}}}^t};\bm{\mu }_{{\text{R}},{\mathbf{z}}}^t,{\mathbf{\Sigma }}_{{\text{R}},{\mathbf{z}}}^t} \right)\mathcal{C}\mathcal{N}\left( {{{\mathbf{z}_{\text{B}}}^t};\bm{\mu }_{{\text{B}},{\mathbf{z}}}^t,{\mathbf{\Sigma }}_{{\text{B}},{\mathbf{z}}}^t} \right),
\end{equation}
where $ {\mathbf{\Sigma }}_{i,{\mathbf{z}}}^t $ and $ \bm{\mu }_{i,{\mathbf{z}}}^t, i\in \{ {\text{R},\text{B}} \} $ can be calculated by
\begin{align}\label{matrix-inversion}
&{\mathbf{\Sigma }}_{i,{\mathbf{z}}}^t = {\left( {{\text{diag}}\left( {{\rho _{i,m,u}}} \right) + \left\langle {{\kappa ^t}} \right\rangle {{\left( {{\mathbf{F}}_i^t\left( {\Delta {{\mathbf{r}}^t}} \right)} \right)}^H}{\mathbf{F}}_i^t\left( {\Delta {{\mathbf{r}}^t}} \right)} \right)^{ - 1}}, \nonumber\\ 
& \qquad \quad  i \in \left\{ {{\text{R}},{\text{B}}} \right\} , 
\end{align}
\begin{align}
\bm{\mu }_{i,{\mathbf{z}}}^t =& \left\langle {{\kappa ^t}} \right\rangle {\mathbf{\Sigma }}_{i,{\mathbf{z}}}^t{\left( {{\mathbf{F}}_i^t\left( {\Delta {{\mathbf{r}}^t}} \right)} \right)^H} \nonumber\\ 
&\resizebox{0.9\hsize}{!}{$
\times \left( {{{\mathbf{y}}^t} - {{\mathbf{\Xi }}^t}\left( {\Delta {\bm{\omega }^{{\text{AoA}}}},\Delta {\bm{\varphi }^{{\text{AoA}}}},\Delta {\bm{\vartheta} ^{{\text{AoA}}}}} \right)\left\langle {{{\mathbf{v}}^t}} \right\rangle } \right),i \in \left\{ {{\text{R}},{\text{B}}} \right\} ,$}
\end{align}
with $ {\mathbf{F}}_{\text{R}}^t\left( {\Delta {{\mathbf{r}}^t}} \right) = \left[ {{{\mathbf{H}}_{r,b}}{\mathbf{\Theta A}}_N^{{\text{LoS}}}\left( 1 \right);...;{{\mathbf{H}}_{r,b}}{\mathbf{\Theta A}}_N^{{\text{LoS}}}\left( G \right)} \right] \in {\mathbb{C}^{KG \times UM}} $ and  $ {\mathbf{F}}_{\text{B}}^t\left( {\Delta {{\mathbf{r}}^t}} \right) = \left[ {{\mathbf{A}}_K^{{\text{LoS}}}\left( 1 \right); ...; }\right. \left. {  {\mathbf{A}}_K^{{\text{LoS}}}\left( G \right)} \right]  \in {\mathbb{C}^{KG \times UM}}$.

\subsubsection{Update Posterior for NLoS Channels $\mathbf{v}^t$}
$ \psi \left( {{{\mathbf{v}}^t}} \right) $'s can be updated by 
\begin{equation}
	\psi \left( {{{\mathbf{v}}^t}} \right) = \mathcal{C}\mathcal{N}\left( {{{\mathbf{v}}^t};\bm{\mu }_{\mathbf{v}}^t,{\mathbf{\Sigma }}_{\mathbf{v}}^t} \right),
\end{equation}
where $ {\mathbf{\Sigma }}_{\mathbf{v}}^t = {\left( {{\text{diag}}\left( {{\gamma _{m,u}}} \right) + \left\langle {{\kappa ^t}} \right\rangle {{\left( {{{\mathbf{\Xi }}^t}} \right)}^H}{{\mathbf{\Xi }}^t}} \right)^{ - 1}} $ and $ \bm{\mu }_{\mathbf{v}}^t = \left\langle {{\kappa ^t}} \right\rangle {\mathbf{\Sigma }}_{\mathbf{v}}^t{\left( {{{\mathbf{\Xi }}^t}} \right)^H}\left( {{{\mathbf{y}}^t} - {{\mathbf{F}}^t}\left( {\Delta {\mathbf{r}}_m^t} \right)\left\langle {{{\mathbf{z}}^t}} \right\rangle } \right) $, with $ \left\langle {\mathbf{z}^t} \right\rangle  = \bm{\mu }_{\mathbf{z}}^t $.

\subsubsection{Update Posteriors for Precision $\bm{\rho}^t$ and $\bm{\gamma}^t$}
The posterior of the precision vector $\bm{\rho}^t$ associated with the LoS channel vector is given by 
\begin{equation}
\psi \left( {{\bm{\rho }^t}} \right) = \prod\limits_{i,m,u} {\Gamma \left( {\rho _{i,m,u}^t;\tilde {\mathfrak{a}}_{i,m,u}^t,\tilde {\mathfrak{b}}_{i,m,u}^t} \right)} ,i \in \left\{ {{\text{R}},{\text{B}}} \right\} ,
\end{equation}
where $ \tilde {\mathfrak{a}}_{i,m,u}^t = \tilde \pi _{m,u}^{\text{L}}\mathfrak{a}_{i,m}^t + \left( {1 - \tilde \pi _{m,u}^{\text{L}}} \right)\bar {\mathfrak{a}}_{i,m}^t + 1 $ and $ \tilde {\mathfrak{b}}_{i,m,u}^t = \tilde \pi _{m,u}^{\text{L}}\mathfrak{b}_{i,m}^t + \left( {1 - \tilde \pi _{m,u}^{\text{L}}} \right)\bar {\mathfrak{b}}_{i,m}^t + {\left| {{{\left[ {\mu _{i,{\mathbf{z}}}^t} \right]}_{m,u}}} \right|^2} + {\left[ {{\mathbf{\Sigma }}_{i,{\mathbf{z}}}^t} \right]_{m,u}} , i\in \{ {\text{R}, \text{B}}\}$. The posterior associated with the off-grid location $ \tilde \pi _{m,u}^{\text{L}} $ is given by (\ref{message-LOS}) shown at the top of the next page,
\begin{figure*}[ht]
\begin{align}\label{message-LOS}
	\tilde \pi _{m,u}^{\text{L}} =& \frac{1}{{{\mathfrak{C}_1}}}\pi _{m,u}^{\text{L}}\exp \left( {\sum\limits_i {\left\{ {\sum\limits_m {\left( {\mathfrak{a}_{i,m}^t - 1} \right)\left( {{\nabla _{\mathfrak{a}_{i,m}^t}}\ln \Gamma \left( {\tilde {\mathfrak{a}}_{i,m,u}^t} \right) - \ln \tilde {\mathfrak{b}}_{i,m,u}^t} \right) - \mathfrak{b}_{i,m}^t\frac{{\tilde {\mathfrak{a}}_{i,m,u}^t}}{{\tilde {\mathfrak{b}}_{i,m,u}^t}}} } \right\}} } \right. \nonumber\\
	&+ \left. { \sum\limits_i {\left\{ {\left( {\bar {\mathfrak{a}}_{i,m}^t - 1} \right)\sum\limits_{u \ne u'} {\left( {{\nabla _{\mathfrak{a}_{i,m}^t}}\ln \Gamma \left( {\tilde {\mathfrak{a}}_{i,m,u'}^t} \right) - \ln \tilde {\mathfrak{b}}_{i,m,u'}^t} \right)}  - \bar {\mathfrak{b}}_{i,m}^t\sum\limits_{u \ne u'} {\frac{{\tilde {\mathfrak{a}}_{i,m,u'}^t}}{{\tilde {\mathfrak{b}}_{i,m,u'}^t}}} } \right\}} } \right),i \in \left\{ {{\text{R}},{\text{B}}} \right\}.
\end{align}
\hrule
\end{figure*}
in which $ {\mathfrak{C}_1} $ is the normalization constant to let $ \sum\nolimits_{u = 1}^U {\tilde \pi _{m,u}^{\text{L}} = 1}  $. Additionally, $ \psi \left( {{\bm{\gamma }^t}} \right) $ is given by
\begin{equation}
\psi \left( {{\bm{\gamma }^t}} \right) = \prod\limits_{m,n} {\Gamma \left( {\gamma _{m,n}^t;\overset{\lower0.1em\hbox{$\smash{\scriptscriptstyle\frown}$}}{\mathfrak{a}} _{m,n}^N,\overset{\lower0.1em\hbox{$\smash{\scriptscriptstyle\frown}$}}{\mathfrak{b}} _{m,n}^N} \right)} ,
\end{equation}
where $\overset{\lower0.1em\hbox{$\smash{\scriptscriptstyle\frown}$}}{\mathfrak{a}} _{m,n}^N = \tilde \pi _{m,n}^{{\text{NL}}}\mathfrak{a}_m^N + \left( {1 - \tilde \pi _{m,n}^{{\text{NL}}}} \right)\bar {\mathfrak{a}}_m^N + 1$ and $ \overset{\lower0.1em\hbox{$\smash{\scriptscriptstyle\frown}$}}{\mathfrak{b}} _{m,n}^N = \tilde \pi _{m,n}^{{\text{NL}}}\mathfrak{b}_m^N + \left( {1 - \tilde \pi _{m,n}^{{\text{NL}}}} \right)\bar {\mathfrak{b}}_m^N + {\left| {\mu _{{\mathbf{v}},m,n}^t} \right|^2} + {\mathbf{\Sigma }}_{{\mathbf{v}},m,n}^t $. 
The posterior associated with the off-grid location $ \tilde \pi _{m,n}^{{\text{NL}}} $ is given by (\ref{message-NLOS}) shown at the top of the next page,
\begin{figure*}
\begin{equation}\label{message-NLOS}
\tilde \pi _{m,n}^{{\text{NL}}} = \frac{1}{{{\mathfrak{C}_2}}}\hat p\left( {c_m^t} \right)\frac{{{{\left( {\mathfrak{b}_m^N} \right)}^{\mathfrak{a}_m^N}}}}{{\Gamma \left( {\mathfrak{a}_m^N} \right)}}\exp \left\{ {\left( {\mathfrak{a}_m^N - 1} \right)\left( {{\nabla _{\mathfrak{a}_m^N}}\ln \Gamma \left( {\mathfrak{a}_m^N} \right) - \ln \mathfrak{b}_m^N} \right) - \mathfrak{b}_m^N\frac{{\overset{\lower0.1em\hbox{$\smash{\scriptscriptstyle\frown}$}}{\mathfrak{a}} _{m,n}^t}}{{\overset{\lower0.1em\hbox{$\smash{\scriptscriptstyle\frown}$}}{\mathfrak{b}} _{m,n}^t}}} \right\}.
\end{equation}
\hrule
\end{figure*}
with the normalization constant $ {\mathfrak{C}_2} $ given by (\ref{constant_C2}) shown at the top of the next page.
\begin{figure*}
\begin{align}\label{constant_C2}
	{\mathfrak{C}_2} =& \hat p\left( {c_m^t} \right)\frac{{{{\left( {\mathfrak{b}_m^N} \right)}^{\mathfrak{a}_m^N}}}}{{\Gamma \left( {\mathfrak{a}_m^N} \right)}}\exp \left\{ {\left( {\mathfrak{a}_m^N - 1} \right)\left( {{\nabla _{\mathfrak{a}_m^N}}\ln \Gamma \left( {\overset{\lower0.1em\hbox{$\smash{\scriptscriptstyle\frown}$}}{\mathfrak{a}} _{m,n}^t} \right) - \ln \overset{\lower0.1em\hbox{$\smash{\scriptscriptstyle\frown}$}}{\mathfrak{b}} _{m,n}^t} \right) - \mathfrak{b}_m^N\frac{{\overset{\lower0.1em\hbox{$\smash{\scriptscriptstyle\frown}$}}{\mathfrak{a}} _{m,n}^t}}{{\overset{\lower0.1em\hbox{$\smash{\scriptscriptstyle\frown}$}}{\mathfrak{b}} _{m,n}^t}}} \right\} \nonumber\\
	&+ \left( {1 - \pi _{m,n}^{{\text{NL}}}} \right)\frac{{{{\left( {\bar {\mathfrak{b}}_m^N} \right)}^{\mathfrak{a}_m^N}}}}{{\Gamma \left( {\bar {\mathfrak{a}}_m^N} \right)}}\exp \left\{ {\left( {\bar {\mathfrak{a}}_m^N - 1} \right)\left( {{\nabla _{\mathfrak{a}_m^N}}\ln \Gamma \left( {\overset{\lower0.1em\hbox{$\smash{\scriptscriptstyle\frown}$}}{\mathfrak{a}} _{m,n}^t} \right) - \ln \overset{\lower0.1em\hbox{$\smash{\scriptscriptstyle\frown}$}}{\mathfrak{b}} _{m,n}^t} \right) - \bar {\mathfrak{b}}_m^N\frac{{\overset{\lower0.1em\hbox{$\smash{\scriptscriptstyle\frown}$}}{\mathfrak{a}} _{m,n}^t}}{{\overset{\lower0.1em\hbox{$\smash{\scriptscriptstyle\frown}$}}{\mathfrak{b}} _{m,n}^t}}} \right\}.
\end{align}
\hrule
\end{figure*}

\subsubsection{Update Posterior for Off-grid Location $\bm{q}^t$}
For the given posterior $ {\tilde \pi _{m,u}^{\text{L}}} $, $\bm{q}^t$'s is updated~by
\begin{equation}\label{update-q}
	\psi \left( {{{\bm{q}}^t}} \right) = \prod\limits_{m} {\sum\limits_u {\tilde \pi _{m,u}^{\text{L}}\delta \left( {q_m^t - u} \right)} } , i \in \left\{ {{\text{R}},{\text{B}}} \right\}.
\end{equation}

\subsubsection{Update Posterior for NLoS Channel Support $\bm{c}^t$}
Using the posterior $ {\tilde \pi _{m,n}^{{\text{NL}}}} $, the NLoS channel support $\bm{c}^t$'s is updated by
\begin{equation}\label{update-c}
\psi \left( {{\bm{c}^t}} \right) = \prod\limits_{m,n} {\tilde \pi _{m,n}^{{\text{NL}}}\delta \left( {c_{m,n}^t - 1} \right) + \left( {1 - \tilde \pi _{m,n}^{{\text{NL}}}} \right)\delta \left( {c_{m,n}^t} \right)} .
\end{equation}

\begin{algorithm}[t]
	\caption{\textbf{D}iverse Dynam\textbf{i}c \textbf{L}ayer Str\textbf{u}ctured \textbf{S}parsity \textbf{S}patial-\textbf{T}emporal \textbf{P}latoon \textbf{L}ocalization (DiLuS-STPL) Algorithm}
	\label{alg}
	\begin{algorithmic}[1]
		\FOR {$ t = 1,2,...,T $}
		\STATE Initialize $\Delta {{\mathbf{r}}^t} = \mathbf{0}$, $ {\left( {\Delta {\bm{\omega }^{{\text{AoA}}}}} \right)^t} = \mathbf{0} $, $ {\left( {\Delta {\bm{\varphi }^{{\text{AoA}}}}} \right)^t} = \mathbf{0} $, $ {\left( {\Delta {\bm{\vartheta} ^{{\text{AoA}}}}} \right)^t} = \mathbf{0} $, hyper-parameters $ \mathfrak{a}_m^t$, $\mathfrak{b}_m^t $, $ \bar {\mathfrak{a}}_m^t$, $\bar {\mathfrak{b}}_m^t $, $ \mathfrak{a}_m^N$, $\mathfrak{b}_m^N $, $ \bar {\mathfrak{a}}_m^N$, $\bar {\mathfrak{b}}_m^N $, $ \mathfrak{a}_m^\kappa$, $\mathfrak{b}_m^\kappa  $,
		the maximum iteration number $r_{\max}$, and the convergence threshold $\epsilon$, $\epsilon '$, $\varepsilon$, $\varepsilon '$.
		\REPEAT
		\STATE {\bf{DiLuS-STPL-E Step:}}
		\STATE {\bf{\% Entity A: Sparse VBI Estimation}}
		\STATE Initialize the distribution according to (\ref{init-q}) and (\ref{init-c}), the message from Entity B passing to Entity A, i.e., $ {\nu _{\chi  \to q_m^t}}\left( {q_m^t} \right)$.
		\REPEAT
		\STATE  Update posteriors $\psi \left( {{{\bf{z}}^t}} \right)$, $\psi \left( {{{\bf{v}}^t}} \right)$, $ \psi \left( {{\bm{\rho }^t}} \right) $, $ \psi \left( {{\bm{\gamma }^t}} \right) $, $\psi \left( {{\bm{q }^t}} \right)$, and $\psi \left( {{\bm{c }^t}} \right)$ in compliance with (\ref{update-z})-(\ref{update-c}) in Sec.~V.~D.
		\UNTIL The convergence criteria of DiLuS-STPL-E Step is met. 
		\STATE Calculate the message from Entity A, i.e., $ {\nu _{{h_{\bm{\rho }}} \to q_m^t}} $, and then send it to Entity B.
		\STATE {\bf{\% Entity B: Forward-backward Message Passing}}
		\STATE 
		Update the message $ {\nu _{\chi  \to q_m^t}}\left( {q_m^t} \right)$, and then send it to Entity A.
		\STATE \textbf{DiLuS-STPL-M Step:} 
		\STATE Construct surrogate functions $ \hat {\mathcal{F}}\left( {{{\bf{\Omega}} ^t};{{\bf{\Omega}} ^{t,\left( r \right)}}} \right) $ in (\ref{surrogate-function}) using the output of Entity A in DiLuS-STPL-E Step, i.e., $ \psi \left( {{\bm{\alpha}}^t} \right) $.
		\STATE Update $ {\bm{\Omega} ^{t,\left( {r + 1} \right)}} $ alternatively according to (\ref{update-Omega}).
		\UNTIL The convergence criteria $ \left\| {{{\bm{\mu }}_{\mathbf{z}}^{t,\left( {r - 1} \right)}} - {{\bm{\mu }}_{\mathbf{z}}^{t,\left( r \right)}}} \right\| \le \epsilon  $, $ \left\| {{{\bm{\mu }}_{\mathbf{v}}^{t,\left( {r - 1} \right)}} - {{\bm{\mu }}_{\mathbf{v}}^{t,\left( r \right)}}} \right\| \le \epsilon ' $, $ \left\| {{{\mathbf{\Sigma }}_{\mathbf{z}}^{t,\left( {r - 1} \right)}} - {{\mathbf{\Sigma }}^{t,\left( r \right)}}} \right\| \le \varepsilon  $, and $ \left\| {{{\mathbf{\Sigma }}_{\mathbf{z}}^{t,\left( {r - 1} \right)}} - {{\mathbf{\Sigma }}_{\mathbf{z}}^{t,\left( r \right)}}} \right\| \le \varepsilon ' $ are met or the maximum iteration number is exceeded.
		\STATE Obtain the optimal posteriors $ {\psi ^*}\left( q_m^t \right) $ and ${\psi ^*}\left( {c_{m,n}^t}\right)$, as well as the parameters ${\bm{\Omega}} ^  {t,*}$. Then, pass both posteriors to the $(t+1)$th slot. 
		\STATE Estimate $\hat q_m^t $ according to $\hat q_m^t = \arg {\max _{{q_t}}}{\psi ^*}\left( {q_m^t} \right)$, and the location estimate in the $t$th slot is given by $ {\mathbf{\hat p}}_m^t = {\mathbf{r}}_{m,\hat q}^t + \Delta {\mathbf{\hat r}}_{m,\hat q}^t $.
		\ENDFOR
	\end{algorithmic}
\end{algorithm}

\subsection{Message Passing from the $(t-1)$th to the $t$th Slot}
The algorithm running in the $t$th slot is dependent on the messages passed from the $(t-1)$th slot, e.g., $ \psi \left( {{\bm{q}^{t - 1}}} \right) $ and $ \psi \left( {{\bm{c}^{t - 1}}} \right) $. Concerning the fact that the approximate posterior $ \psi \left( {{\bm{q}^{t - 1}}} \right) $ can be taken from the forward-backward message passing in (\ref{approx-previous-slot}) (as is the same for $ \psi \left( {{\bm{c}^{t - 1}}} \right) $), the priors $ {\left( { \pi _{m,u}^{\text{L}}} \right)^{t }} $ and $ \hat p\left( {{\bm{c}^t}} \right) $ can be obtained in light of $ \pi _{m,u}^{\text{L}} = {\nu _{\chi  \to q_m^t}}\left( {q_m^t} \right) $ and $ \hat p\left( {{\bm{c}^t}} \right) $, the former of which incorporates the messages over the forward-backward propagation while the latter of which is derived from (\ref{init-c}). Then, the posteriors  $ {\left( {\tilde \pi _{m,u}^{\text{L}}} \right)^{t }} $ and $ {\left( {\tilde \pi _{m,n}^{{\text{NL}}}} \right)^{t}} $ can be updated according to (\ref{message-LOS})~and~(\ref{message-NLOS}). With the closed-form updates in (\ref{update-q}) and (\ref{update-c}), the message passing across two adjacent slots can thus be achieved. In summary, the proposed DiLuS-STPL algorithm is summarized in Algorithm~1.

\subsection{Computational Complexity and Practical Implementation}
This subsection discusses the computational complexity and offers a glimpse into the practical implementation of the proposed DiLuS-STPL algorithm. Following the overall flow of DiLuS-STPL outlined in Algorithm~\ref{alg}, the main computational burden is dominated by the updates of the posteriors $\psi \left( {{{\bf{z}}^t}} \right)$, $\psi \left( {{{\bf{v}}^t}} \right)$, $ \psi \left( {{\bm{\rho }^t}} \right) $, $ \psi \left( {{\bm{\gamma }^t}} \right) $, $\psi \left( {{\bm{q }^t}} \right)$, and $\psi \left( {{\bm{c }^t}} \right)$ in Step~8 using (\ref{update-z})-(\ref{update-c}), respectively. More precisely, the computational complexity of the matrix inversion operation in (\ref{matrix-inversion}) is $ \mathcal{O}\left( {{U^3}{M^3}} \right) $, along with the total number of multiplications for updating $\psi \left( {{{\bf{z}}^t}} \right)$ being $ \mathcal{O}\left( {{K^3}{G^3} + 3UM + 2KG{U^2}{M^2} + 2{K^2}{G^2}UM + {K^2}{G^2}} \right) $. Thus, the complexity order can be given by $ \mathcal{O}\left( {KG{U^2}{M^2}} \right) $, since it typically holds that  $KG \ll UM $. Regarding its practical implementation, the low-latency requirements of the considered RIS-aided vehicle platoons may not be satisfied when running DiLuS-STPL based on a regular desktop computer with a PYTHON implementation. It is just what we use to provide a theoretical guideline. This issue can be effectively addressed by leveraging the powerful and parallel computing power of the cloud, in which the BS first sends the received signals ${\bf{y}}^t$ to the cloud, and then the cloud runs the DiLuS-STPL algorithm to attain the exact locations of $M$ VUEs based on the received signals ${\bf{y}}^t$. Therefore, the low-latency requirements of the practical system can be met due to the large-bandwidth and high-rate fronthaul links between the BS and the cloud~\cite{C-RAN}.

\section{Simulation Results}\label{sec-Simulation}
\subsection{Simulation Configuration}
A vehicle platoon system consisting of $M$ VUEs is customized according to the generation procedure in \cite{VP-1}. To be specific, the spatial correlation of the $M$ VUEs positioned on a road is modeled using a sampled Gamma distribution as presented in (\ref{location-gamma}) with the shape and scale parameters given by $\varpi = 2$ and $\lambda = 1.5$. Concerning that the inter-VUE distance is a random variable due to the fluctuation in the speed control, it is assumed that the speed of each VUE follows a complex Gaussian distribution $ \mathcal{C}\mathcal{N}\left( {\bar v,{{\bar \sigma }^2}} \right) $ with $ \bar v =  - 18{\text{ m/s}} $ as its mean and $\bar \sigma  = 8{\text{ m/s}}$ as its standard deviation. In this case, the temporal probability, i.e., $h_{T,m}^t$, is given by
\begin{align}
h_{T,m}^t &\triangleq p\left( {q_m^t|q_m^{t - 1}} \right) \nonumber\\
&= {\left( {\frac{{\bar \sigma \Delta t}}{{\Delta L}}\sqrt {2\pi } } \right)^{ - 1}}\exp \left( { - \frac{1}{2}{{\left( {\frac{{q_m^t - \left( {q_m^{t - 1} + \frac{{\bar v\Delta t}}{{\Delta L}}} \right)}}{{\bar \sigma \Delta t/\Delta L}}} \right)}^2}} \right),
\end{align}
in which $\Delta L =  \left\| {{\mathbf{r}}_{m,u}^t - {\mathbf{r}}_{m,u - 1}^t} \right\| $  denotes the grid length and $ \Delta t $ is the slot interval. The BS and the RIS are positioned on each side of the road, both of which are oriented parallel to the road. The three-dimensional coordinates of the BS and the RIS in meters are $ \left[ {50,100,25} \right] $ and $ \left[ {150,0,25} \right] $, respectively. The considered system operates at 7~GHz. Furthermore, the exploited primary performance metric for location estimates is the root-mean-square-error (RMSE), i.e., $ \sqrt {\frac{1}{{TSM}}\sum\nolimits_{t = 1}^T {\sum\nolimits_{s = 1}^S {\sum\nolimits_{m = 1}^M {{{\left\| {{\mathbf{\hat r}}_m^t - {\mathbf{r}}_m^t} \right\|}^2}} } } }  $, where $ S $ is the number of random channel realizations generated in a slot, and thus the estimate output for each slot $t$ have been averaged over these $S$ samples. The number of observation slots for the system is $ T = 100 $. Other system parameters are set as follows if not otherwise specified: $M=4$, $K=16$, $N=256$, $\zeta_{m,b} = 3$, $\zeta_{m,r}= 2.5$, $\zeta_{r,b}= 2$ $ \Delta L = 1{\text{ m}} $, $ \Delta t = 0.1 {\text{ s}} $,  $S=200$, $\sigma^2=-100\ \text{dBm}$, and $\widetilde{\varrho}_{m,n}=0.3$.

\begin{figure}
		\centering
		\includegraphics[width=0.5\textwidth]{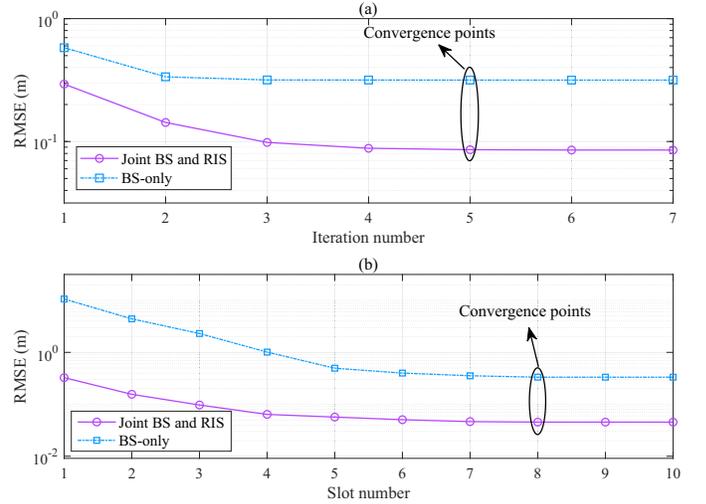}
		\caption{Convergence behaviors of the proposed DiLuS-STPL algorithm.} \label{Converge1}
\end{figure}
\begin{figure}
		\centering
		\includegraphics[width=0.5\textwidth]{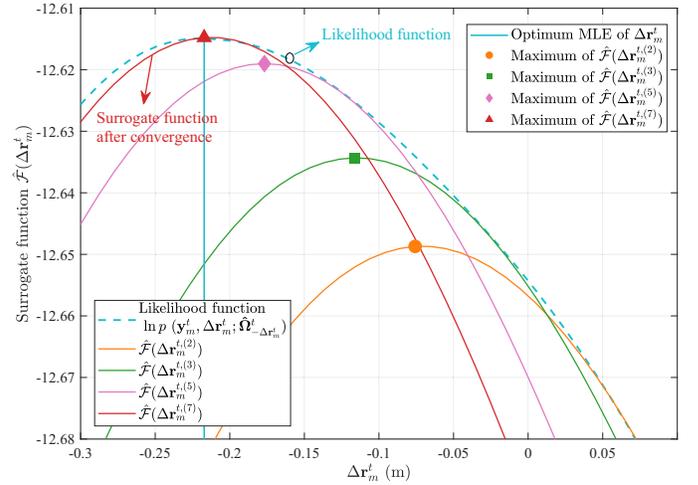}
		\caption{Surrogate function $\hat{\mathcal{F}}\left( {\Delta \mathbf{r}_m^t}\right) $ versus the estimated location $ {\Delta \mathbf{r}_m^t} $ during the convergence procedure in a slot.} \label{Converge2}
\end{figure}

\subsection{Convergence Behavior}
In Fig.~\ref{Converge1}, we show the convergence behavior of the proposed algorithm by plotting the RMSE curves associated with the estimated location ${\mathbf{\hat r}}_m^t$. Specifically, it is observed in Fig.~\ref{Converge1}(a) that the RMSE performance decreases rapidly with the iteration number, and ultimately converges to a value within around 5 iterations in a slot. Fig.~\ref{Converge1}(b) demonstrates that the RMSE performance tends to stabilize over time, and DiLuS-STPL eventually converges to a better performance. In the first slot, DiLuS-STPL performs poor due primarily to the unavailability of any priori provided. With an increase of the slot, DiLuS-STPL takes advantage of the temporal correlations and the priori delivered by the previous slot for an appropriate parameter adjustment. Upon achieving the performance saturation, the location tracking process progressively transitions into a steady state.

In order to confirm that the surrogate function adopted in~(\ref{surrogate-function}) typically provides a safe approximation of the likelihood function in (\ref{pro-ML}), Fig.~\ref{Converge2} depicts the variation of surrogate function $\hat{\mathcal{F}}\left( {\Delta \mathbf{r}_m^t}\right) $ versus the estimated location $ \left( {\Delta \mathbf{r}_m^t}\right) $ during the convergence procedure in a slot. We are primarily concerned  with the impact introduced by the estimated off-grid location, and thus it is assumed here that other angle-associated offsets, e.g.,  $ \Delta {{\mathbf{r}}^t}$, ${\left( {\Delta {\bm{\omega }^{{\text{AoA}}}}} \right)^t}$, ${\left( {\Delta {\bm{\varphi }^{{\text{AoA}}}}} \right)^t}$, and ${\left( {\Delta {\bm{\vartheta} ^{{\text{AoA}}}}} \right)^t} $, are perfectly available during the estimation. It can be intuitively observed that the surrogate function, as the iteration number increases, approaches the original likelihood more closely. As DiLuS-STPL converges in the seventh iteration, the estimated location $ \left( {\Delta \mathbf{r}_m^t}\right) $ (also the maximum of $\hat{\mathcal{F}}\left( {\Delta \mathbf{r}_m^{t,\left( 7 \right) }}\right) $) almost coincides with an optimum MLE, indicating that our theoretical analyses are in excellent agreement with simulation results.

\begin{figure}
		\centering
		\includegraphics[width=0.5\textwidth]{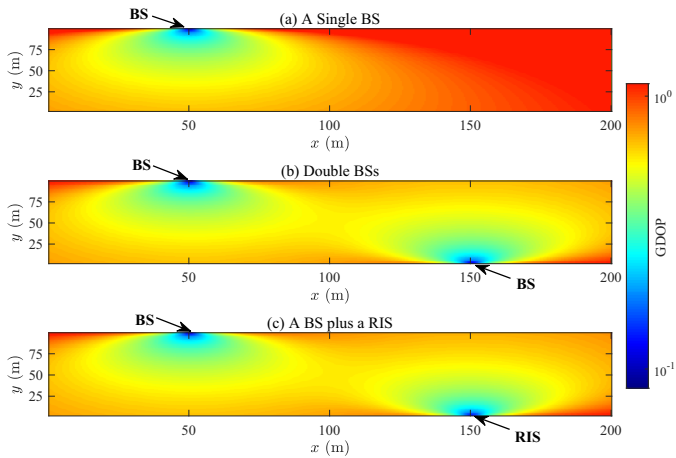}
		\caption{GDOP performance versus the varying VUE location in different points of the road when $K = 4$, $N = 256$.} \label{GDOP1}
\end{figure}

\subsection{The Impact of RIS on Localization}
\subsubsection{GDOP Performance Evaluation}
Fig.~\ref{GDOP1} plots the GDOP performance versus the varying VUE location in different points of the road, in which the location tracking relies exclusively upon a single BS positioned at $ \left[ { 50,100, 25} \right] (\rm{m}) $ in (a); a pair of BSs operate cooperatively for location tracking in (b), with their positions given by $ \left[ { 50,100,25} \right] (\rm{m}) $ and $ \left[ {150,0,25} \right] (\rm{m})$, respectively; a BS and a RIS operate in tandem in (c), with their positions given by $ \left[ { 50,100,25} \right] (\rm{m}) $ and $ \left[ {150,0,25} \right] (\rm{m})$, respectively. Firstly, as observed, the achieved GDOP is dependent on the relative VUE location with respect to the BS and RIS. Only when the VUE approaches the BS or RIS will a lower estimation error be achieved, whereas the localization precision deteriorates immediately as soon as the VUE moves further away from them. Given that the uniform grid is fixed, the angular resolution ${\left\| {\varphi _m^{{\text{AoA}}}\left( {{\mathbf{r}}_{m,u}^t + \Delta {\mathbf{r}}_{m,u}^t} \right) - \varphi _m^{{\text{AoA}}}\left( {{\mathbf{r}}_{m,u - 1}^t + \Delta {\mathbf{r}}_{m,u - 1}^t} \right)} \right\|_F}$ associated with the off-grid location declines with an increased $\frac{{{\mathbf{r}}_{m,u}^t}}{{{h_v}}}$ as $u$ grows larger, which leads to the unfavorable column-non-orthogonality in the LoS array response $ {\mathbf{A}}_K^{{\text{LoS}}}\left( {\varphi _m^{{\text{AoA}}}\left( {{\mathbf{r}}_{m,u}^t + \Delta {\mathbf{r}}_m^t} \right)} \right) $, eroding the precision of estimates. Secondly, the dual BSs cooperating for localization provides the best-case bound by virtue of their powerful signal processing capabilities. Intriguingly, we notice that the case of the BS and RIS operating in tandem exhibits the almost comparable localization performance with respect to GDOP, with only marginal precision loss spotted on the side of the RIS. This can be attributed to the following facts. The passive nature of the RIS regulates its reflective behavior in response to an incident signal. The localization signal, in this context, may entail attenuation due to the double fading effect \cite{chen-jsac}, leading to a drop in propagation signal-to-noise ratio (SNR) and an increase in estimate error. Despite this, no noticeable performance loss is spotted, which reveals RIS's great potentials for enabling high-accuracy localization owing to its cost-effective implementations in practice.

\begin{figure}
	\centering
	\includegraphics[width=0.5\textwidth]{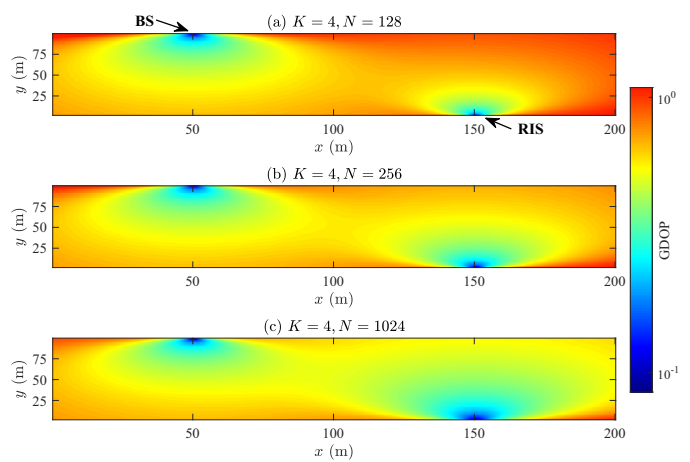}
	\caption{GDOP performance achieved by a BS plus a RIS for different antenna/element configurations.} \label{GDOP2}
\end{figure}

\subsubsection{Localization Capabilities: BS vs. RIS}
One may argue that the double-BS configuration could supplant the combination of a BS and a RIS, albeit their comparable localization performance demonstrated in Fig.~\ref{GDOP1}. To further justify this, we would like to go deeper into the rationale of employing a BS and RIS for localization based upon the simulation results in Fig.~\ref{GDOP2}. Specifically, the GDOP ratio $  {\text{GDO}}{{\text{P}}_{{\text{BS}}}}/{\text{GDO}}{{\text{P}}_{{\text{RIS}}}}  $ presented in (\ref{GDOP_ratio}) motivates us to explore just their individual localization capabilities in a neat manner, instead of more complicated GDOP forms, e.g., ${\text{GDOP}}_{\text{BS\&RIS}}$ and ${\text{GDOP}}_{\text{BS\&BS}}$. Thus, we are interested in the case of a BS and a RIS functioning in tandem with different antenna/element configurations. The BS and the RIS are placed at $ \left[ { 50,100,25} \right] (\rm{m}) $ and $ \left[ {150,0,25} \right] (\rm{m})$, respectively. It is intuitive in Fig.~\ref{GDOP2} that an increase in $ N $ boosts RIS's localization capability in terms of the GDOP performance. This occurrence is in good agreement with the fact drawn in (\ref{GDOP_ratio}), since the ratio $  {\text{GDO}}{{\text{P}}_{{\text{BS}}}}/{\text{GDO}}{{\text{P}}_{{\text{RIS}}}}  $ is proportionate to $ N_h $ and  $ N_v $. A larger $N$ has the potential to compensate for the multiplicative fading nature inherent to the cascaded channels, thus achieving favorable localization performance. Therefore, we would prefer to implement a BS plus a RIS rather than the double-BS configuration, due to the extremely low implementation cost of a RIS (about \$ 7000 \cite{price-RIS}) in comparison to the high cost of a 5G BS (about \$ 23,000 \cite{price-BS}). Additionally, concerning the passive nature of the RIS, additional interference cannot be introduced to the users in a single cell.

\begin{figure*}[t]
	\centering
	\includegraphics[width=1.0\textwidth]{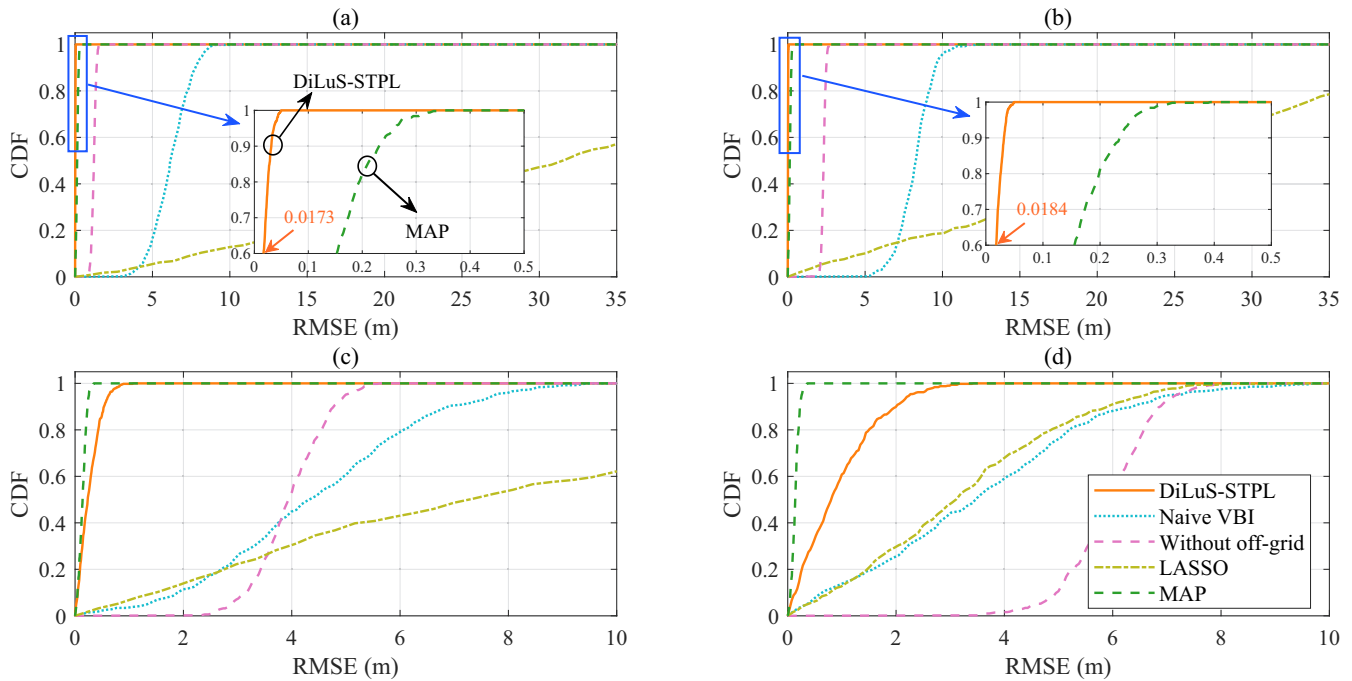}
	\caption{CDF of the RMSE for various grid resolutions. (a)~The grid length is $ \Delta L = 0.5 \text{ m} $, and the number of grids is $U=200$. (b)~The grid length is $ \Delta L = 1 \text{ m} $, and the number of grids is $U=100$. (c)~The grid length is $ \Delta L = 5 \text{ m} $, and the number of grids is $U=40$. (d) The grid length is $ \Delta L = 20 \text{ m} $, and the number of grids is $U=10$.} \label{CDF}
\end{figure*}

\subsection{Performance of DiLuS-STPL}
For comparison, we employ the following several benchmark schemes:
\begin{itemize}
	\item Naive VBI \cite{SPM-2008}: It is assumed that the naive VBI algorithm employs i.i.d. priors for both LoS and NLoS channels without considering the diverse sparsity structures presented in DiLuS.
	
	\item MAP \cite{book-BMTT}: The fundamental of MAP is a Bayesian filter employed to approximate the posteriors of the corresponding parameters, i.e.,  ${{\mathbf{r}}^t}$, ${\left( {{\bm{\omega }^{{\text{AoA}}}}} \right)^t}$, ${\left( {{\bm{\varphi }^{{\text{AoA}}}}} \right)^t}$, and ${\left( {{\bm{\vartheta} ^{{\text{AoA}}}}} \right)^t}$, whose extremum points can be determined by a brute-force-kind approach.

	\item  LASSO \cite{LASSO}: The LASSO algorithm is a widely used CS-based sparse signal recovery technique that directly filters the non-zero entries in a sparse vector by a soft or hard threshold.
		
	\item  Without off-grid: The proposed DiLuS-STPL is utilized for location tracking by employing the sparse priors presented in DiLuS without factoring the off-grid basis, i.e., the offsets $ \Delta {{\mathbf{r}}^t}$, ${\left( {\Delta {\bm{\omega }^{{\text{AoA}}}}} \right)^t}$, ${\left( {\Delta {\bm{\varphi }^{{\text{AoA}}}}} \right)^t}$, and ${\left( {\Delta {\bm{\vartheta} ^{{\text{AoA}}}}} \right)^t} $ are set to be zero.
\end{itemize}

\subsubsection{Impact of Grid Resolution}
We present in Fig.~\ref{CDF} the CDF of RMSE under several schemes for different grid resolutions. It can be observed from Fig.~\ref{CDF} that our proposed DiLuS-STPL achieves highly desirable RMSE performance for all grid resolution cases and even outperforms that of MAP for the instances illustrated in Figs.~\ref{CDF}(a)\&(b), i.e., the grid length $\Delta L =  \left\| {{\mathbf{r}}_{m,u}^t - {\mathbf{r}}_{m,u - 1}^t} \right\| $ being both 0.5~m and 1~m, as well as the number of grids being both $ U=200 $ and $ U=100 $. This is attributed to the smoothness and convexity of the surrogate function used in DiluS-STPL, as well as the beneficial sparse priors provided by DiLuS, significantly facilitating the posterior approximation and parameter updates. Furthermore, as transpired in Figs.~\ref{CDF}(a)\&(b), a finer-grained gird resolution delivers fairly slight improvement in RMSE performance but adversely doubles the search space, which might make no sense in the practical algorithm design since we are always striving for a critical compromise between the precision and computational complexity. By comparison, due to the absence of these properties in MAP, its performance is bounded by the grid-by-grid search, in which the grid resolution becomes the bottleneck for achieving a higher level of localization precision. Furthermore, concerning that the approximate posteriors in MAP are independent of the grid length, RMSE somehow does not vary with the grid length using MAP approaches. MAP's use of a brute-force-kind approach allows it still to perform excellently even at coarse grid resolutions, but at the sacrifice of an extremely high computational complexity. 
Additionally, a somewhat counterintuitive occurrence in Fig.~\ref{CDF} is that the RMSE performance tends to be improved as the grid resolution lowers for the LASSO scheme. This is because LASSO is highly dependent on the column-orthogonality in the sensing matrix~\cite{AngDom-11}. A refined grid length results in marginal differences between the neighboring columns in the AoA array response, i.e., $ {\mathbf{A}}_K^{{\text{LoS}}}\left( {\varphi _m^{{\text{AoA}}}\left( {{\mathbf{r}}_{m,u}^t + \Delta {\mathbf{r}}_m^t} \right)} \right) $, hence rendering the undesirable correlations that may be detrimental to the accuracy of grid estimates.

\begin{figure}
	\subfigure[]{
		\begin{minipage}[t]{0.5\textwidth}
			\centering
			\includegraphics[width=\textwidth]{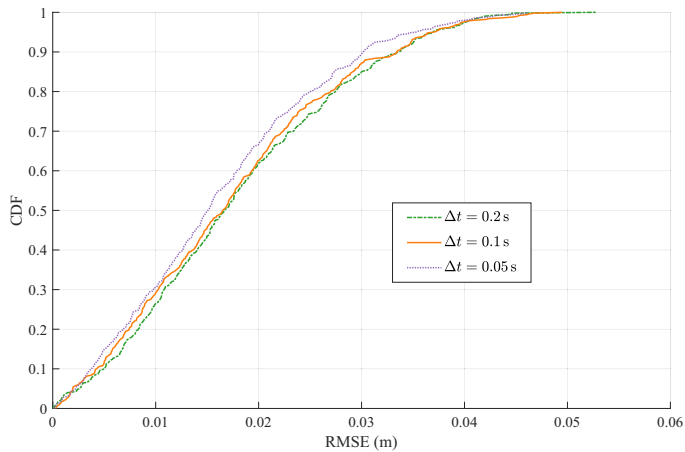}
			\label{TemporalCDF}
	\end{minipage}}
	\subfigure[]{
		\begin{minipage}[t]{0.50\textwidth}
			\centering
			\includegraphics[width=\textwidth]{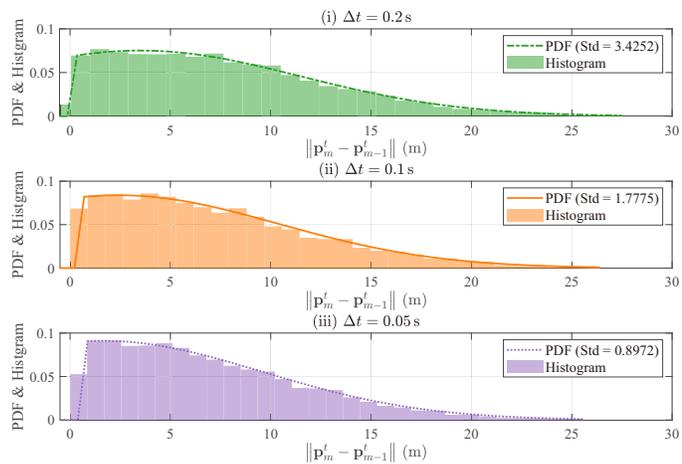} 
			\label{TruncatedGaussian}
	\end{minipage}}
	\caption{(a) CDF of the RMSE for various temporal resolutions. (b) Truncated Gaussian distributions with various variances associated with different $\Delta t$.}\label{TemporalResolution}
\end{figure}

\subsubsection{Impact of Temporal Resolution}

Fig.~\ref{TemporalResolution} shows the CDF of the RMSE performance achieved by DiLuS-STPL with different temporal resolutions, i.e., $\Delta t = 0.05 \, \text{s}$, $\Delta t = 0.1 \,  \text{s}$, and $\Delta t = 0.2 \, \text{s}$. As transpired in Fig.~\ref{TemporalCDF}, the RMSE performance might be slightly improved if the slot interval is shortened. Despite this, their differences in regard to the three temporal resolutions are marginal at best and, in practice, may be ignored. Furthermore, Fig.~\ref{TruncatedGaussian} depicts an analysis of the possible causes for this slight improvement. In particular, the samples for each temporal resolution follow a truncated Gaussian distribution whose variance is independent on $\Delta t$. A finer-grained $\Delta t$ induces a smaller variance, thus allowing for a more subtle vibration of RMSE throughout the whole observation procedure.

\subsubsection{Impact of the Number of RIS Elements}
\begin{figure}
		\centering
		\includegraphics[width=0.5\textwidth]{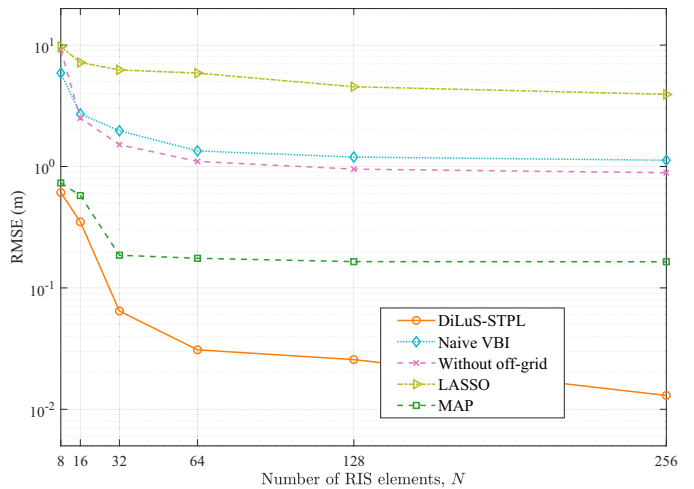}
		\caption{RMSE performance versus the number of RIS elements.} \label{RMSEvsN}
\end{figure}

In Fig.~\ref{RMSEvsN}, the RMSE performance is plotted for different numbers of RIS elements $N$, when the number of BS antennas is fixed with $K=16$. It is evident that RMSE performance obtained by all schemes improves with an increased $N$ due to the enhancement in angular resolution. As $N$ grows to a certain magnitude, neither the naive VBI nor the MAP scheme demonstrates additional precision gains with respect to RMSE. This is due primarily to the absence of a tailored design for capturing additional NLoS priors brought by the cross-slot temporal correlations, as well as the inability to capture the location priors. Furthermore, we note that the RMSE performance of DiLuS-STPL without off-grid levels off as $N$ grows because the estimation error is in general bounded by the grid resolution as portrayed in Fig.~\ref{CDF}. This can also be attributed to the fact that DiLuS-STPL without off-grid tends to converge around the grid length. Therefore, we evince that the proposed DiLuS-STPL is capable of harnessing the potential benefits brought by the improved angular resolution, since it encapsulates various sparsity structures associated with the LoS/NLoS channels and the spatial-temporal correlations inherent to the vehicle platoon, in an attempt to recursively identify exact priors for high-accuracy location tracking.

\begin{figure}
		\centering
		\includegraphics[width=0.5\textwidth]{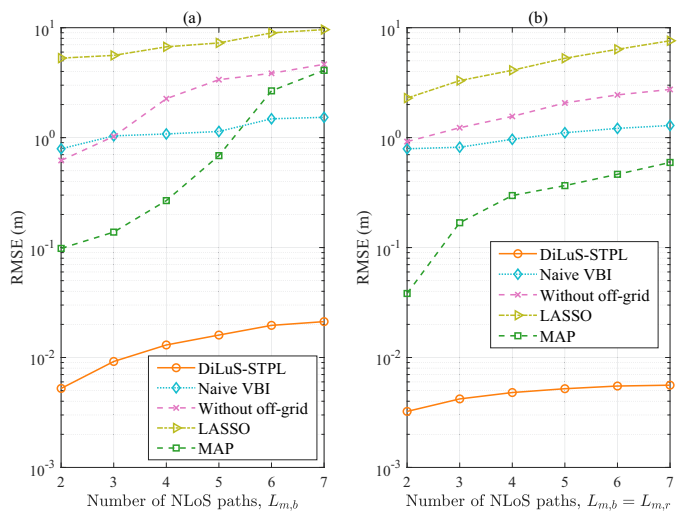}
		\caption{RMSE performance versus the number of NLoS paths. (a) Location tracking relies exclusively upon the BS. (b) The BS and the RIS jointly track VUE's location.} \label{RMSEvsPath}
\end{figure}

\subsubsection{Impact of the Number of NLoS Paths}
Fig.~\ref{RMSEvsPath} investigates the RMSE performance versus the varying number of NLoS paths in cases of only the BS being active as well as the BS and RIS functioning in tandem. It is assumed for simplicity that the direct and the cascaded channels hold the same number of NLoS paths, i.e., $L_{m,b} = L_{m,r}$. Firstly, RMSE is intuitively observed to somewhat deteriorate with an increasing number of NLoS paths, particularly when only the BS is responsible for the localization. The increased number of NLoS paths makes it prohibitive to distinguish the LoS paths that carries useful information for location tracking from the NLoS paths, referred to as the NLoS-induced misleading effect, thus eroding the accuracy of localization. Fortunately, the favorable LoS paths constructed by the RIS enriches the localization-associated information, along with the significantly improved accuracy by limiting the NLoS-induced misleading effect towards the LoS paths. Secondly, despite the fluctuating number of NLoS paths, the proposed DiLuS-STPL algorithm levels off even in the presence of an increase in number of NLoS paths, showing its great robustness. Besides, several benchmarks are able to mitigate the NLoS-induced misleading effect while identifying the location-associated LoS channels with the aid of the RIS. This facilitates the precise and robust localization despite the intricate dynamics of the vehicular environment.

\section{Conclusion}\label{sec-conclusion}

In this work, we have examined the RIS as an extension of the BS's localization capability as it provides additional location priors and location-associated LoS paths. By fully exploiting the diverse sparsities that are available in the LoS/NLoS channels, a DiLuS framework is customized for encapsulating the sparse priors of corresponding channels, followed by a MAP-form location tracking problem that is highly intractable due to the uncertain sensing matrix and coupled latent variables associated with the BS and RIS. To resolve this hindrance,  the DiLuS-STPL algorithm is developed by recursively achieving updates of the parameters, i.e., sparse LoS/NLoS channel vectors, off-grid location and angle offsets. Our simulation results have verified the significant potential of the RIS to enable high-accuracy localization. We evince through the GDOP metric that location tracking using a BS and a RIS may arrive at the comparable precision performance obtained by the two individual BSs, which is an exciting discovery due to the cost-effective attributes of the RIS in practice. Additionally, the proposed DiLuS-STPL algorithm has the potential to mitigate the NLoS-induced misleading effect while attempting to identify the LoS channels in the presence of a fluctuating number of NLoS paths, thus revealing the critical robustness for high-accuracy localization in vehicle platoons.

\appendix[Derivation in (\ref{GDOP_RIS})]\label{Appendix_GDOP}
Derivation is likewise performed by first achieving the outcome for the ULA case, and then extending it to the UPA case. As for the cascaded channel ${{\mathbf{H}}_{r,b}}{\mathbf{\Theta h}}_{m,r}^t$, the entry ${\mu _{{\text{B}},k}}$ in (\ref{FIM_omega}) can be replaced by ${\mu _{{\text{R}},k}}$ that represents the $(k,n)$th entry of ${{\mathbf{H}}_{r,b}}{\mathbf{\Theta h}}_{m,r}^t$. With the AoA $\varphi _m^{{\text{AoA}}}$, the FIM can be written as 
\begin{align}\label{FIM_varphi_1}
&{\text{FIM}}\left( {\varphi _m^{{\text{AoA}}}} \right) \nonumber\\ &= \frac{2}{{{\sigma ^2}}}\operatorname{Re} \left\{ {{\nabla _{\varphi _m^{{\text{AoA}}}}}{{\left( {{{\mathbf{H}}_{r,b}}{\mathbf{\Theta h}}_{m,r}^t} \right)}^H}{\nabla _{\varphi _m^{{\text{AoA}}}}}\left( {{{\mathbf{H}}_{r,b}}{\mathbf{\Theta h}}_{m,r}^t} \right)} \right\} \nonumber\\
 &= \frac{2}{{{\sigma ^2}}}\operatorname{Re} \left\{ {{\nabla _{\varphi _m^{{\text{AoA}}}}}{{\left( {{\mathbf{h}}_{m,r}^t} \right)}^H} \left(  {{\mathbf{\Theta }}^H}{\mathbf{H}}_{r,b}^H{{\mathbf{H}}_{r,b}}{\mathbf{\Theta }} \right)  {\nabla _{\varphi _m^{{\text{AoA}}}}}{\mathbf{h}}_{m,r}^t} \right\}.
\end{align}
By letting  $ {\widetilde {\mathbf{H}}_{r,b}} = {{\mathbf{H}}_{r,b}}{\mathbf{\Theta }} $, the auto-correlation matrix can be given by $ \widetilde {\mathbf{H}}_{r,b}^H{\widetilde {\mathbf{H}}_{r,b}} $  whose diagonal entries tend to zero due to its sparsity in the angular-domain representation. The $\left( {n,n} \right)$th entry of $ \widetilde {\mathbf{H}}_{r,b}^H{\widetilde {\mathbf{H}}_{r,b}} $  can thus be approximated as $ {\left[ {\widetilde {\mathbf{H}}_{r,b}^H{{\widetilde {\mathbf{H}}}_{r,b}}} \right]_{n,n}} \approx {\left| {{\text{P}}{{\text{L}}_{r,b}}} \right|^2}K $. Furthermore, (\ref{FIM_varphi_1}) can be recast~to 
\begin{equation}\label{FIM_varphi}
{\text{FIM}}\left( {\varphi _m^{{\text{AoA}}}} \right) = {\left| {{\text{P}}{{\text{L}}_{r,b}}} \right|^2}K \cdot \frac{2}{{{\sigma ^2}}}\operatorname{Re} \left\{ {\sum\limits_{n = 1}^N {\frac{{\partial \left( {{\mathbf{h}}_{m,r}^t} \right)_n^*}}{{\partial \varphi _m^{{\text{AoA}}}}}\frac{{\partial {{\left( {{\mathbf{h}}_{m,r}^t} \right)}_n}}}{{\partial \varphi _m^{{\text{AoA}}}}}} } \right\}.
\end{equation}
We observe that the structural distinction between (\ref{FIM_omega}) and (\ref{FIM_varphi}) resides in a cascaded channel-induced extra term. Thus, by following the flow from (\ref{CRLB_ULA}) to (\ref{GDOP_UPA}), we can obtain $ {\text{GDO}}{{\text{P}}_{{\text{RIS}}}}$ in~(\ref{GDOP_RIS}).


%

%
%
%


\ifCLASSOPTIONcaptionsoff
  \newpage
\fi

\bibliographystyle{IEEEtran}
\bibliography{ref_RIS_LOC}

\begin{IEEEbiography}[{\includegraphics[width=1in,height=1.25in,clip,keepaspectratio]{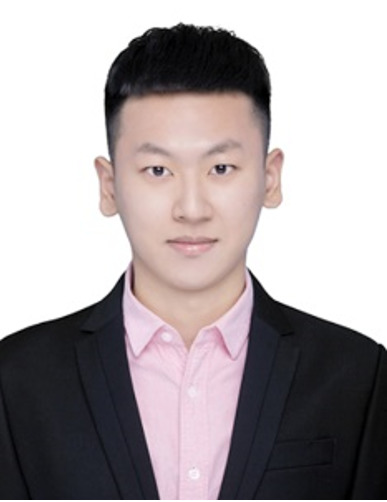}}]{Yuanbin Chen}
	received the B.S. degree in communications engineering from the Beijing Jiaotong University, Beijing, China, in 2019. He is currently pursuing the Ph.D. degree in information and communication systems with the State Key Laboratory of Networking and Switching Technology, Beijing University of Posts and Telecommunications. His current research interests are in the area of reconfigurable intelligent surface (RIS), vehicle-to-everything (V2X), and radio resource management (RRM) in future wireless networks. He was the recipient of the National Scholarship in 2020 and 2022.
\end{IEEEbiography}

\begin{IEEEbiography}[{\includegraphics[width=1in,height=1.25in,clip,keepaspectratio]{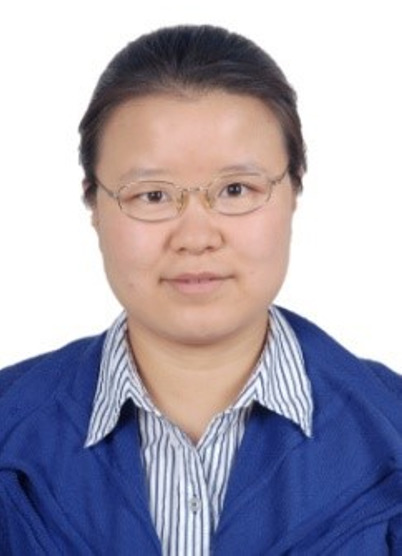}}]{Ying Wang}
	(IEEE Member) received the Ph.D. degree in circuits and systems from the Beijing University of Posts and Telecommunications (BUPT), Beijing, China, in 2003. 
	In 2004, she was invited to work as a Visiting Researcher with the Communications Research Laboratory (renamed NiCT from 2004), Yokosuka, Japan. She was a Research Associate with the University of Hong Kong, Hong Kong, in 2005. She is currently a Professor with BUPT and the Director of the Radio Resource Management Laboratory, Wireless Technology Innovation Institute, BUPT. Her research interests are in the area of the cooperative and cognitive systems, radio resource management, and mobility management in 5G systems. She is active in standardization activities of 3GPP and ITU. She took part in performance evaluation work of the Chinese Evaluation Group, as a Representative of BUPT. She was a recipient of first prizes of the Scientific and Technological Progress Award by the China Institute of Communications in 2006 and 2009, respectively, and a second prize of the National Scientific and Technological Progress Award in 2008. She was also selected in the New Star Program of Beijing Science and Technology Committee and the New Century Excellent Talents in University, Ministry of Education, in 2007 and 2009, respectively. She has authored over 100 papers in international journals and conferences proceedings.
\end{IEEEbiography}

\begin{IEEEbiography}[{\includegraphics[width=1in,height=1.25in,clip,keepaspectratio]{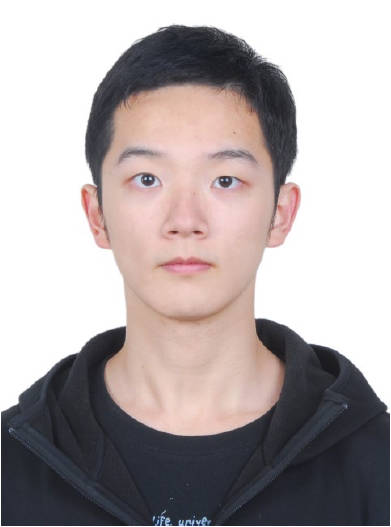}}]{Xufeng Guo} received the B.E. degree in communications engineering from the Beijing University of Posts and Telecommunications, Beijing, China, in 2021. He is currently pursuing the Ph.D. degree in information and communication systems with the State Key Laboratory of Networking and Switching Technology, Beijing University of Posts and Telecommunications. His current research interests are in the area of reconfigurable intelligent surface (RIS), extremely large-scale multiple-input-multiple-output (XL-MIMO), and signal processing techniques in future wireless networks.
\end{IEEEbiography}

\begin{IEEEbiography}[{\includegraphics[width=1in,height=1.25in,clip,keepaspectratio]{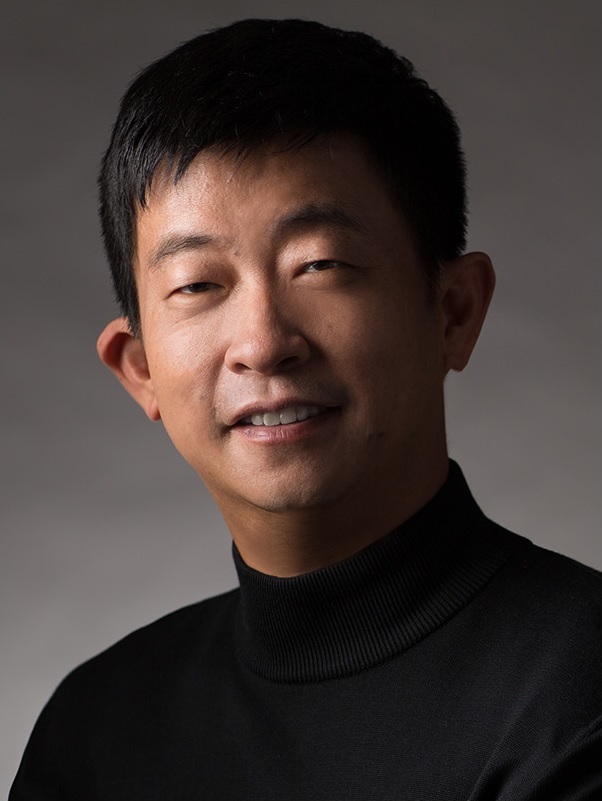}}]{Zhu Han}
	(S’01–M’04-SM’09-F’14) received the B.S. degree in electronic engineering from Tsinghua University, in 1997, and the M.S. and Ph.D. degrees in electrical and computer engineering from the University of Maryland, College Park, in 1999 and 2003, respectively. 
	
	From 2000 to 2002, he was an R\&D Engineer of JDSU, Germantown, Maryland. From 2003 to 2006, he was a Research Associate at the University of Maryland. From 2006 to 2008, he was an assistant professor at Boise State University, Idaho. Currently, he is a John and Rebecca Moores Professor in the Electrical and Computer Engineering Department as well as in the Computer Science Department at the University of Houston, Texas. Dr. Han’s main research targets on the novel game-theory related concepts critical to enabling efficient and distributive use of wireless networks with limited resources. His other research interests include wireless resource allocation and management, wireless communications and networking, quantum computing, data science, smart grid, security and privacy.  Dr. Han received an NSF Career Award in 2010, the Fred W. Ellersick Prize of the IEEE Communication Society in 2011, the EURASIP Best Paper Award for the Journal on Advances in Signal Processing in 2015, IEEE Leonard G. Abraham Prize in the field of Communications Systems (best paper award in IEEE JSAC) in 2016, and several best paper awards in IEEE conferences. Dr. Han was an IEEE Communications Society Distinguished Lecturer from 2015-2018, AAAS fellow since 2019, and ACM distinguished Member since 2019. Dr. Han is a 1\% highly cited researcher since 2017 according to Web of Science. Dr. Han is also the winner of the 2021 IEEE Kiyo Tomiyasu Award, for outstanding early to mid-career contributions to technologies holding the promise of innovative applications, with the following citation: ``for contributions to game theory and distributed management of autonomous communication networks."
\end{IEEEbiography}

\begin{IEEEbiography}[{\includegraphics[width=1in,height=1.25in,clip,keepaspectratio]{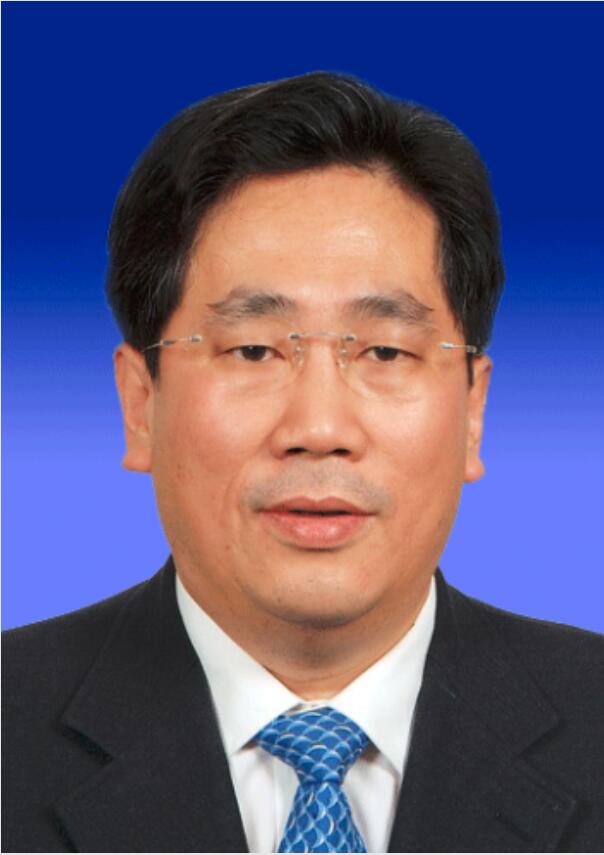}}]{Ping Zhang} (IEEE Fellow) is a Professor with the School of Information and Communication Engineering, Beijing University of Posts and Telecommunications, the Director of the State Key Laboratory of Networking and Switching Technology, a member of IMT-2020 (5G) Experts Panel, and a member of Experts Panel for China’s 6G Development. He served as a Chief Scientist of National Basic Research Program (973 Program), an Expert in information technology division of National High-Tech Research and Development Program (863 Program), and a member of Consultant Committee on International Cooperation of National Natural Science Foundation of China. His research interests mainly focus on wireless communications.
\end{IEEEbiography}

\end{document}